\documentclass[aps,showpacs,prc,floatfix,onecolumn]{revtex4} 
\usepackage{graphicx}
\usepackage{amssymb}
\usepackage{latexsym} 
\usepackage{epsfig}
\newcommand{\dii}{{\mathrm d}}

\newcommand{\vew}{\vec{w}}
\newcommand{\vet}{\vec{t}}

\newcommand{\veq}{\vec{q}}
\newcommand{\veQ}{\vec{Q}}
\newcommand{\vetau}{\vec{\tau}}

\newcommand{\ii}{{\rm i}}
\newcommand{\e}{{\rm e}}
\newcommand{\p}{{\rm p}}
\newcommand{\Lab}{\lambda_B}
\newcommand{\Las}{\lambda_S}
\newcommand{\Laq}{\lambda_Q}
\newcommand{\ranGC}{\rangle_{\mathrm{GC}}}
\newcommand{\GC}{{\mathrm{GC}}}

\newcommand{\ch}{{\mathrm{ch}}}

\newcommand{\zjl}{z_{j(1)}}
\newcommand{\zkl}{z_{k(1)}}
\newcommand{\zjn}{z_{j(n_j)}}
\newcommand{\zkn}{z_{k(n_k)}}
\newcommand{\Omj}{\Omega^{(j)}}
\newcommand{\Omk}{\Omega^{(k)}}
\newcommand{\Omjk}{\Omega^{(jk)}}
\newcommand{\hF}{\overline{F}}

\def\Nj{{\{N_j\}}}

\def\hpartj{\{h_{n_j}\}}
\begin{document}

\title{Multiplicity fluctuations in the hadron gas with exact conservation laws}

\author{Francesco Becattini}
\email{becattini@fi.infn.it}
\affiliation{Universit\`a di Firenze and INFN Sezione di Firenze,
Via G. Sansone 1, I-50019 Sesto Fiorentino, Firenze}
\author{Antti Ker\"anen}
\email{antti.keranen@oulu.fi}
\affiliation{Universit\`a di Firenze and INFN Sezione di Firenze,
Via G. Sansone 1, I-50019 Sesto Fiorentino, Firenze}
\author{Lorenzo Ferroni}
\email{ferroni@fi.infn.it}
\affiliation{Universit\`a di Firenze and INFN Sezione di Firenze,
Via G. Sansone 1, I-50019 Sesto Fiorentino, Firenze}
\author{Tommaso Gabbriellini}
\email{gabbriellini@fi.infn.it}
\affiliation{Universit\`a di Firenze and INFN Sezione di Firenze,
Via G. Sansone 1, I-50019 Sesto Fiorentino, Firenze}

\begin{abstract}
The study of fluctuations of particle multiplicities in relativistic heavy-ion 
reactions has drawn much attention in recent years, because they have been 
proposed as a probe for underlying dynamics and possible formation of quark-gluon 
plasma. Thus, it is of uttermost importance 
to describe the baseline of statistical fluctuations in the hadron gas phase 
in a correct way. We have performed a comprehensive study of multiplicity 
distributions in the full ideal hadron-resonance gas in different ensembles, 
namely grand-canonical, canonical and microcanonical, using two different 
methods: asymptotic expansions and full Monte Carlo simulations. The method
based on asymptotic expansion allows a quick numerical calculation of dispersions
in the hadron gas with three conserved charges at primary hadron level, while 
the Monte-Carlo simulation is suitable to study the effect of resonance decays. 
Even though mean multiplicities converge to the same values, major differences 
in fluctuations for these ensembles  persist in the thermodynamic limit, as 
pointed out in recent studies. We observe that this difference is ultimately 
related to the non-additivity of the variances in the ensembles with exact 
conservation of extensive quantities.    
\end{abstract}
\maketitle

\section{Introduction}

The statistical hadronization model has proven to be a very effective tool in describing 
average particle multiplicities in high energy heavy ion reactions \cite{vari} 
as well as in elementary particle reactions \cite{elem}. On the other 
hand, in this model it would be possible to calculate also fluctuations of particle 
multiplicities once the status of the hadronizing sources ({\em clusters} or 
{\em fireballs}) in terms of volume, mass, momentum and charges were known. 
Multiplicity and charge fluctuations have been indeed proposed to be a good 
discriminating tool between quark-gluon plasma and hadron gas \cite{jeon, heinz}
provided that they survive the phase transition and the hadronic system freezes out 
in a non-equilibrium situation. However, in order to properly assess the 
discriminating power of such observables, one should firstly calculate fluctuations in 
a hadron gas by including all ``trivial" effects, such as conservation laws, quantum 
statistics, resonance decays, kinematical cuts etc. The effects of conservation laws
on fluctuations in thermal ensembles have been firstly addressed, in the perspective 
of heavy ion collisions, in ref.~\cite{raja}. More
recently, it has been pointed out \cite{gore1,gore2} that in the canonical ensemble 
(CE) with exact conservation of charges, scaled second moment (scaled variance) of 
the multiplicity distribution of any particle does not converge to the corresponding
GC value even in the thermodynamic limit, unlike the mean \cite{cleymans,kerabeca}. 
This was fairly understood among experts in statistical mechanics \cite{touchpr}, 
but probably it has been shown explicitely for the canonical relativistic gas for the
first time in refs.~\cite{gore1,gore2}. Further deviations from the GC limit 
were found in the case of exact energy and energy-momentum conservation in the 
microcanonical ensemble (MCE) \cite{becaferro2,gore3}. Since in a heavy ion collision 
conservation of charges must be fulfilled, the difference between CE and GCE might 
have some impact on the estimated size of fluctuations in a statistical model. 

The calculations performed in these recent studies \cite{gore1,gore2,gore3,turko} were
mainly concerned with simplified cases, such as pion gas and pion-nucleon gas. 
In the present work, we address the fluctuations in the {\em general} multi-species
hadron gas including all resonances up to 1.8 GeV mass and carrying three additive
charges, that is baryon number $B$, strangeness $S$ and electric charge $Q$, in 
the CE and MCE. We give first a general formulation of the problem, and derive novel 
fluctuation formulae for the large volume 
limit in the CE and MCE. We also discuss the problem of the inequivalence between
GCE and CE in the thermodynamic limit for scaled variance and we show that the 
ultimate reason thereof is the conceptual difference between additivity and extensivity
\cite{touch}: while particle multiplicities are additive and extensive in both GCE and 
CE, variances are extensive (i.e. proportional to the volume) but they are non-additive 
in the CE, so that the scaled variance turns out to be a {\em pseudo-intensive} 
quantity (according to the definition proposed in ref.~\cite{touch}).

The paper is organized as follows: in the next section we address the large volume 
limit of fluctuations and correlations in the canonical ensemble by the means of 
asymptotic expansions. Monte-Carlo calculations are described and compared to analytical
ones in Sect.~3. In Sect.~4 we address the large volume limit of fluctuations in
the microcanonical ensemble. In Sect.~5 we discuss the difference between thermodynamic
limits of fluctuations in the different ensembles and relate them to fundamental 
properties of the scaled variance. In Sect.~6 we address the problem of measuring
fluctuations of charged particles and calculate different proposed measures in 
the different ensembles. The results are summarized in Sect.~7.
 
\section{Asymptotic fluctuations in the canonical ensemble}

Following refs.~\cite{gore1,gore2}, we describe fluctuations by means of the {\em 
scaled variance} of a multiplicity distribution: 
\begin{equation} \label{omega}
\omega = \frac{\langle N^2 \rangle -\langle N \rangle^2}{\langle N \rangle}.
\end{equation}
where $N$ is meant to be the multiplicity of any hadron species, primary or final
(i.e. after resonance decays) or the sum of an arbitrary number of hadron species
(e.g. all negatively chraged). This is a finite quantity in the infinite volume limit 
because the difference between $\langle N^2 \rangle$ and $\langle N \rangle^2$ depends 
linearly on the volume for large volumes. It is worth reminding that, if quantum 
statistics is neglected, the multiplicity distribution of any {\em primary} hadron is
a Poisson, thus $\omega=1$.

For the sake of simplicity, we will first keep our discussion at the level of the classical 
Maxwell-Boltzmann (MB) statistics. Indeed, none of our arguments is affected by 
this approximation, and at the end we will give the proper generalization to quantum
statistics and discuss the corrections. In this framework, using the one-particle 
partition function:
\begin{equation}\label{zeta}
 \zjl = (2J_j+1)\frac{V}{(2\pi)^3} 
 \int \dii^3 \p \; \exp \left[ -\sqrt{\p^2+m^2_j} \right]
\end{equation}
and the fugacity $\lambda_j$ for each particle species $j$, the grand-canonical 
partition function can be written as:
\begin{equation} \label{eq:ZGC}
 Z_{\GC}(\{\lambda_j\}) = \prod_j \sum_{N_j=0}^\infty \frac{1}{N_j!}
\left(\zjl \lambda_j\right)^{N_j}.
\end{equation}
and, consequently, multiplicities of different species are uncorrelated and 
Poissonianly distributed:
\begin{equation}\label{poisson}
P_{\GC}(N_j) =  \frac{1}{N_j!} \langle N_j \rangle^{N_j} \e^{-\langle N_j \rangle}.
\end{equation}
Since the sum of random Poisson variables is still Poisson, this also holds 
for any given subset of particles, e.g. negative hadrons or baryons.

In the canonical ensemble, the partition function does not factorize into one-species
expressions because of the constraint of fixed charges. Let us consider a hadron gas 
with three abelian charges, i.e. baryon number $B$, strangeness $S$ and electric charge 
$Q$. In the following, we will denote by $\vec{Q} = (Q_1,Q_2,Q_3) = (B,S,Q)$ a vector 
with components these charges and by $\veq_j = (q_{1,j},q_{2,j},q_{3,j}) = (b_j,s_j,q_j)$
the vector of charges of the $j^{\rm th}$ hadron species \footnote{For the sake of clarity, 
it is worth stressing the difference between $q_{i,j}$, which is the $i\,{\rm th}$ 
charge of the hadron species $j$ and $q_j$, which stands for its {\em electric charge}. 
Likewise, whilst $Q_i$ stands for the net $i\,{\rm th}$ charge of the system, $Q$ is
its net electric charge throughout the paper}. 
The canonical partition function with charges $\veQ$ can be written as: 
\begin{equation} \label{projection}
Z_{\veQ} = \left[ \prod_{i=1}^{3} \frac{1}{2\pi} 
 \int_0^{2\pi}\dii\phi_i \; \e^{-\ii Q_i\phi_i}\right] Z_{GC}(\{\lambda_j\}),
\end{equation}
where Wick-rotated fugacities $\lambda_j=\exp[\ii \sum_i q_{i,j} \phi_i]$ are 
introduced in the grand-canonical partition function $Z_{GC}$. By setting 
$w_i = \exp[\ii \phi_i]$, we may write Eq.~(\ref{projection}) as a triple integral 
over the unitary circle in the complex $w$ plane:
\begin{equation}\label{canpart}
Z_{\vec{Q}} = \frac{1}{(2\pi\ii)^3} \oint \dii w_B \oint \dii w_S \oint \dii w_Q
\, w_B^{-B-1} w_S^{-S-1} w_Q^{-Q-1}
\exp\!\left[ \sum_j \zjl w_B^{b_j} w_S^{s_j} w_Q^{q_j} \right].
\end{equation}
The first and second moments of multiplicity distributions of a set $h$ of hadron
species can be calculated by inserting a suitable fictitious fugacity in the function
$Z_{GC}$, i.e. replacing $\lambda_j$ with $\lambda_h \lambda_j$ in 
Eq.~(\ref{projection}) if $j \in h$ and taking the derivatives with respect to 
$\lambda_h$ \cite{becaheinz}:
\begin{eqnarray} \label{moments}
\langle N_h \rangle &= \displaystyle{ 
\frac{1}{Z_{\vec{Q}}} \frac{\partial Z_{\vec{Q}}}{\partial \lambda_h}
\Bigg|_{\lambda_h=1}
}
&= \sum_{j\in h} \zjl \frac{Z_{\vec{Q}-\veq_j}}{Z_{\vec{Q}}}\\
\langle N_h^2 \rangle &= \displaystyle{
\frac{1}{Z_{\vec{Q}}} \left[\frac{\partial}{\partial\lambda_h}\left(\lambda_h
\frac{\partial Z_{\vec{Q}}} {\partial \lambda_h}\right)\right]_{\lambda_h=1}
}
&= \sum_{j\in h} \zjl \frac{Z_{\vec{Q}-\veq_j}}{Z_{\vec{Q}}}
+  \sum_{j,k\in h} \zjl \zkl \frac{Z_{\vec{Q}-\veq_j-\veq_k}}
{Z_{\vec{Q}}}
\end{eqnarray}
Using these, the scaled variance can be written as the sum of a Poissonian term,
i.e. 1, and a canonical correction term:
\begin{equation} \label{MBomega} 
\omega_h = 1 + \frac{\sum_{j \in h} \langle N_j \rangle \sum_{ k \in h} 
 \zkl \left( \displaystyle{ 
 \frac{Z_{\veQ-\veq_k-\veq_j}}{Z_{\veQ-\veq_j}}- \frac{Z_{\veQ-\veq_k}}{Z_{\veQ}}} 
 \right)}{\sum_{j \in h} \langle N_j \rangle}
\end{equation}
Therefore, in the canonical ensemble, the quantities appearing in the expressions
of the moments of the multiplicity distributions are the canonical partition functions 
calculated for the difference between total charges and charges of hadrons, like
$Z_{\veQ-\veq_j}$ and $Z_{\veQ-\veq_j-\veq_k}$.
The quantity within brackets in the equation above vanishes in the thermodynamic 
limit $V \to \infty$, as $Z_{\vec Q} \sim \exp(-{\vec \mu} \cdot {\vec Q}/T)$ (see
Eq.~(\ref{limgc}) below), ${\vec \mu}$ being the vector of chemical potentials 
corresponding to the conserved charges ${\vec Q}$. However, the factor 
$\langle N_j \rangle \zkl$ is proportional to $V^2$ and, if the difference between
brackets has terms proportional to $1/V$, they could give a finite contribution
to $\omega_h$ in the thermodynamic limit. In fact, we will show that if this limit 
is properly taken, this is generally the case. Let us first rewrite $Z_{\veQ-\veq_j}$
by using Eq.~(\ref{canpart}) as:
\begin{eqnarray}\label{partx}
Z_{\veQ-\veq_j} \!\!&=& \!\! \frac{1}{(2\pi\ii)^3} \oint \dii w_B \oint \dii w_S \oint 
\dii w_Q \ w_B^{b_j-1}w_S^{s_j-1} w_Q^{q_j-1} \exp \left[-B\ln w_B-S\ln w_S-Q\ln w_Q
+\sum_k \zkl w_B^{b_k}w_S^{s_k}w_Q^{q_k}\right] \nonumber \\
&\equiv& \frac{1}{(2\pi\ii)^3} \oint \dii w_B \oint \dii w_S \oint \dii w_Q \,
g(\vec{w}) \exp [V f(\vec{w})]. 
\end{eqnarray}
where: 
\begin{eqnarray}
  \label{eq:g}
  g(\vew) &=& w_B^{b_j-1}w_S^{s_j-1}w_Q^{q_j-1} \\
  \label{eq:f}
  f(\vew) &=& -\rho_B \ln w_B - \rho_S \ln w_S - \rho_Q \ln w_Q +
  \sum_k \frac{\zkl}{V} w_B^{b_k}w_S^{s_k}w_Q^{q_k}
\end{eqnarray}
$\rho_B = B/V$, $\rho_S = S/V$, $\rho_Q=Q/V$ being the baryon, strangeness and 
electric charge densities respectively. 

The thermodynamic limit of the moments will be calculated by using an asymptotic
saddle-point expansion of the above integral in the {\em large} parameter $V$ 
keeping $\rho_B,\rho_S,\rho_Q$ fixed. Note that the function $f$ in Eq.~(\ref{eq:f})
will then be independent of the volume because so is $\zkl/V$ according to 
Eq.~(\ref{zeta}). 
On the other hand, $g(\vec{w})$ does not depend on large parameters because hadron 
charges are limited to few units. Furthermore, with this choice, the
function $f(\vec{w})$ has a saddle point $\vec{w}_0$ determined by the condition 
${\partial f(\vec{w})}/{\partial w_i}=0$ which is independent of the hadron charges
$\veq_j$ in Eq.~(\ref{partx}). It can be shown that the saddle point coincides 
with physical fugacities in the grand-canonical ensemble, i.e. $\vec{w}_0 = 
(\Lab,\Las,\Laq)$ \cite{becahi3}. For instance:
\begin{eqnarray}\label{saddle}
 && \frac{\partial f}{\partial w_Q} = -\frac{\rho_Q}{w_Q} + \sum_k q_k \frac{\zkl}{V} 
 w_B^{q_k}w_S^{s_k}w_Q^{q_k-1} = 0 \nonumber \\
 &\Rightarrow& \sum_k q_k \zkl w_{0B}^{b_k}w_{0S}^{s_k}w_{0Q}^{q_k} = Q
\end{eqnarray}
and similarly for $w_B$ and $w_S$. The last equation, along with the corresponding
equations for $B$ and $S$, manifestly show that the solution $\vec{w}_0$ must coincide 
with the fugacities in the GCE, as in this case $\zkl w_B^{b_k}w_S^{s_k}w_Q^{q_k} = 
\langle N_k \rangle_{\GC}$. The leading factor in the saddle-point expansion of the
canonical partition function $Z_{\vec Q}$ will then be (see also Eq.~(\ref{core})):
\begin{eqnarray}\label{limgc}
 \exp[V f(\vew_0)] 
 &=& \exp \left[ -B \ln \lambda_B - S \ln \lambda_S - Q \ln \lambda_Q + 
 \sum_k \zkl \lambda_B^{b_k} \lambda_S^{s_k} \lambda_Q^{q_k} \right] \nonumber \\ 
 &=& \exp \left[-{\vec \mu} \cdot {\vec Q}/T \right] \exp \left[\sum_k \zkl 
 \e^{{\vec \mu} \cdot {\vec q}_k/T} \right] = 
 \exp \left[ -{\vec \mu} \cdot {\vec Q}/T \right] Z_{\GC} 
\end{eqnarray}
i.e. proportional to the grand-canonical partition function itself.

We will now proceed to calculate the large volume limits of the scaled variance 
$\omega$ in the pion gas and in the full hadron gas. The saddle-point asymptotic 
expansions of the canonical partition 
functions, for different set of charges, will not be derived in detail in the 
following. Their general expression is discussed and quoted in Appendix A. 

\subsection{Illustrative example: classical pion gas}
\label{subs:pion}

The fluctuations in the canonical pion gas in Boltzmann statistics have been recently 
studied by Gorenstein {\em et. al.} \cite{gore1,gore2}. Even though the pion gas 
(including $\pi^+$ and $\pi^-$, $\pi^0$ is irrelevant in this context) has 
only one conserved abelian charge, we apply the more general approach of saddle-point 
expansions, which is better suited in case of several conserved charges. This serves 
as a good introduction to the full hadron gas case. 

The partition function of the pion gas for a total charge $Q$ can be obtained 
at the order ${\mathcal O}(V^{-3/2})$ by using the general formula of the saddle-point 
expansion (\ref{core}) quoted in Appendix A, with $\nu = V$, $g(w)=1/w$, $f(w)= 
- \rho_Q \ln w + z_{\pi}/V (w +1/w)$ and the saddle-point $w_0 = \Laq$ being the 
grand-canonical fugacity:
\begin{eqnarray}
  Z^\pi_Q &=& \frac{Z_{\GC}^\pi}{\Laq^{Q}} \sqrt{\frac{1}{2 \pi V f''(\Laq)}}
  \left[\frac{1}{\Laq} + \frac{1}{V} \left(\frac{\gamma(\Laq)}{\Laq}-
  \frac{\alpha(\Laq)}{\Laq^2} - \frac{1}{\Laq^3 f''(\Laq)} \right)  
   + \mathcal{O}\left(V^{-2}\right) \right]  \nonumber \\
  &=& \frac{Z_{\GC}^\pi}{\Laq^{Q+1}} \sqrt{\frac{1}{2 \pi V f''(\Laq)}}
  \left[ 1 + \frac{1}{V} \left({\gamma(\Laq)}-\frac{\alpha(\Laq)}{\Laq} 
   - \frac{1}{\Laq^2 f''(\Laq)} \right)  
   + \mathcal{O}\left(V^{-2}\right) \right]
\end{eqnarray}   
where $\alpha(\Laq)$ and $\gamma(\Laq)$ are constants dependent on derivatives of the 
function $f$ beyond second order in the saddle-point. 
Now $f''(\Laq) = \Laq^{-2}\left[\rho_Q + 2z_\pi/V \Laq \right]$ and defining 
$C \equiv Z_{\GC}^\pi/\Laq^{Q+1} \sqrt{1/ 2\pi V f''(\Laq)}$ which is a constant factor
independent of the pion charges, the previous equation turns into:
\begin{equation} \label{piZ1}
  Z^\pi_Q = C \left(1 + \frac{\gamma(\Laq)}{V}-\frac{\alpha(\Laq)}{V \Laq} 
   - \frac{1}{Q + 2\langle \pi^- \rangle_\GC} \right) + \mathcal{O}(V^{-2})
\end{equation}
where we have used $\langle\pi^-\rangle_\GC = z_\pi/\Laq$, i.e. the expression of
the grand-canonical mean multiplicity of negative pions, in the limit of Boltzmann 
statistics. Similarly, the asymptotic 
saddle-point expansion of the canonical partition function for a charge $Q-Q_i$ 
can be obtained by taking $g(\vew) = w_Q^{Q_j-1}$ in Eq.~(\ref{core}):
\begin{eqnarray} \label{piZ2}
 Z^{\pi}_{Q-Q_j} &=& \frac{Z_{\GC}^\pi}{\Laq^{Q+1}} \sqrt{\frac{1}{2 \pi V f''(\Laq)}} 
 \Laq^{q_j} \left\{ 1 + \frac{1}{V} \left[ \gamma(\Laq) + 
  (q_j-1) \frac{\alpha(\Laq)}{\Laq} - \frac{1}{2} (q_j-1)(q_j-2) 
 \frac{1}{\Laq^2 f''(\Laq)} \right] + \mathcal{O}(V^{-2})\right\} \nonumber \\
 &=& C \Laq^{q_j} \left[ 1 + \frac{\gamma(\Laq)}{V} + (q_j-1) \frac{\alpha(\Laq)}
 {V \Laq} - \frac{1}{2} (q_j-1)(q_j-2) \frac{1}{Q + 2\langle \pi^- \rangle_\GC} \right] + 
 \mathcal{O}(V^{-2}),
\end{eqnarray}   
Note that Eq.~(\ref{piZ1}) is, as expected, a special case of Eq.~(\ref{piZ2}) with 
$q_j = 0$. From Eq.~(\ref{moments}) and (\ref{piZ1}),(\ref{piZ2}) we deduce 
that the mean multiplicity of pions converges to the grand-canonical value in
the thermodynamic limit $V \to \infty$:
\begin{equation}\label{eq:pisar}
\langle \pi^\pm \rangle = z_\pi \frac{Z^{\pi}_{Q \mp 1}}{Z^{\pi}_{Q}}=
 z_\pi \Laq^{\pm 1} + \mathcal{O}\left(V^{-1}\right) 
 = \langle \pi^\pm \ranGC + \mathcal{O}\left(V^{-1}\right).
\end{equation}
On the other hand, this is not the case for the scaled variance $\omega$. In fact,
the part in the parentheses in Eq.~(\ref{MBomega}) yields a term of order 
$\mathcal{O}(V^{-1})$ stemming from the difference between the terms proportional
to $1/(Q+2\langle \pi^- \rangle_\GC)$ in Eqs.~(\ref{piZ1}),(\ref{piZ2}). On the
other hand, the two terms proportional to $\alpha$ and $\gamma(\Laq)$, cancel out at the 
order $\mathcal{O}(V^{-1})$ in the difference between $Z_{Q-Q_j-Q_k}/Z_{Q-Q_j}$ and 
$Z_{Q-Q_k}/Z_{Q}$. Specifically: 
\begin{equation}
\frac{Z^{\pi}_{Q-Q_j-Q_k}} {Z^{\pi}_{Q-Q_j}} - \frac{Z^{\pi}_{Q-Q_k}}{Z^{\pi}_Q} = 
-\Laq^{q_k}q_k q_j\frac{1}{Q+2\langle\pi \ranGC} + \mathcal{O}\left(V^{-2}\right). 
\end{equation} 
Inserting this and (\ref{eq:pisar}) into Eq.~(\ref{MBomega}), and writing 
$Q=\langle \pi^+ \ranGC - \langle \pi^- \ranGC$, we obtain the final expression of
scaled variances for both net charge and total particle number distributions:
\begin{eqnarray} \label{eq:pipmomega}
\lim_{V \to \infty} \omega_\pm &=& 1-\frac{\langle \pi^\pm \ranGC}
{\langle \pi^+ \ranGC + \langle \pi^- \ranGC} \\
                 \label{eq:pichomega}
\lim_{V\rightarrow\infty}\omega_\ch &=& 1-\left( 
\frac{\langle \pi^+ \ranGC - \langle \pi^- \ranGC}
{\langle \pi^+ \ranGC + \langle \pi^- \ranGC}\right)^2.
\end{eqnarray} 
These nice and simple expressions reduce to the grand-canonical (Poissonian) limit 
1 only in special cases, and never for all three quantities under the same conditions.
From Eq.~(\ref{eq:pipmomega}) it can be seen that the Poisson value of 1 for the 
scaled variance is recovered in the thermodynamical limit for negatives if 
$\langle \pi^+ \ranGC  \gg \langle \pi^- \ranGC$ (and vice-versa), i.e. when the 
net charge of the system is largely positive and the oppositely charged particles 
become a small pollution; conversely, the scaled variance of the positive particles 
almost vanishes because $\langle \pi^+ \ranGC \simeq Q$, which must not fluctuate.

\subsection{Full hadron gas} 
\label{subs:general}

We now turn to the general case of multi-hadron gas carrying total charges 
$\{Q_l\} = (B,S,Q)$. The asymptotic expansion of the canonical partition function 
involves diagonalization of the Hessian matrix of the function $f$ defined in 
Eqs~(\ref{partx}) and (\ref{eq:f}) (see Appendix A):
\begin{equation}\label{hessian}
 H_{l n}(\vew_0) = \frac{\partial^2 f(\vew)}
{\partial w_{Q_l}\partial w_{Q_n}}\Bigg|_{\vew_0} = 
\frac{1}{V \lambda_{Q_l}\lambda_{Q_n}} \left[ Q_l \delta_{ln} + 
\sum_j q_{l,j} \left( q_{n,j}-\delta_{ln}\right) \langle N_j \rangle_{\mathrm{GC}}
\right\}.
\end{equation}
where $q_{l,j}$ is the $l^{\rm th}$ charge of the $j^{\rm th}$ hadron species.
Because this is real and symmetric, the diagonalization is achieved through an 
orthogonal matrix ${\sf A}$:
\begin{equation}
 {\sf H}' = {\mathrm{diag}}(h_1,h_2,h_3) = {\sf A} {\sf H} {\sf A}^T
\end{equation}
where $h_l$ are the Hessian eigenvalues. It should be emphasized that the hessian 
matrix ${\sf H}$, as well as its eigenvalues and the diagonalizing matrix ${\sf A}$, 
is not dependent on the volume, just like the function $f$ in Eq.~(\ref{eq:f}). By 
using the general expression (\ref{core}) quoted in Appendix A with $f$ in 
Eq.~(\ref{eq:f}) and $g$ in Eq.~(\ref{eq:g}), it can be proved that 
the asymptotic expansions in the volume $V$ of the partition function $Z_{\veQ-\veq_j}$ 
reads:
\begin{eqnarray}\label{sar1}
 Z_{\veQ-\veq_j} &=& \frac{Z_{\rm GC}}{\Lab^{B+1}\Las^{S+1}\Laq^{Q+1}} 
 \sqrt{\frac{1}{(2\pi)^3 V^3 \det {\sf H}}} \Lab^{b_j}\Las^{s_j}\Laq^{q_j}
 \Bigg\{1 + \frac{1}{V}\left[ -\frac{1}{2\Lab^2}(b_j-1)(b_j-2)\left(\sum_{i=1}^{3}
\frac{A_{i1}A_{i1}}{h_i}\right)\right.\nonumber\\
&& - \frac{1}{2\Las^2}(s_j-1)(s_j-2) \left(\sum_{i=1}^{3}
 \frac{A_{i2}A_{i2}} {h_i}\right) 
- \frac{1}{2\Laq^2}(q_j-1)(q_j-2) \left(\sum_{i=1}^{3} 
\frac{A_{i3}A_{i3}} {h_i}\right) \nonumber \\
&& -\frac{1}{\Lab\Las} (b_j-1)(s_j-1)\left(\sum_{i=1}^{3}\frac{A_{i1}
A_{i2}}{h_i}\right) -\frac{1}{\Lab\Laq}(b_j-1)(q_j-1)
\left(\sum_{i=1}^{3}\frac{A_{i1}A_{i3}}{h_i}\right) \nonumber \\
&&\left.\left. - \frac{1}{\Laq\Las}(s_j-1)(q_j-1)
\left(\sum_{i=1}^{3}\frac{A_{i2}A_{i3}}{h_i}\right) + \gamma
+ \frac{\alpha_B(b_j -1)}{\Lab} + \frac{\alpha_S(s_j-1)}{\Las} + 
 \frac{\alpha_Q(q_j-1)}{\Laq}\right]\right\} + \mathcal{O}\left(V^{-2}\right). 
\end{eqnarray}
A straightforward consequence of Eq.~(\ref{sar1}) is:
\begin{equation}
 \lim_{V \to \infty} \frac{Z_{\veQ-\veq_j}}{Z_{\veQ}} = 
 \Lab^{b_j}\Las^{s_j}\Laq^{q_j} \equiv \lambda_j,
\end{equation}
so that the usual grand-canonical expression is recovered for the average 
multiplicity in the large volume limit:
\begin{equation}\label{nGC}
  \lim_{V \to \infty} \zjl \frac{Z_{\veQ-\veq_j}}{Z_{\veQ}} =
  \zjl \Lab^{b_j}\Las^{s_j}\Laq^{q_j} = \langle N_j \rangle_\GC
\end{equation}  
Furthermore, denoting:
$$M_{ln} = \sum_{i=1}^{3}\frac{A_{il}A_{in}}{h_i}.$$
and using Eq.~(\ref{sar1}), it can be straightforwardly proved that:
\begin{eqnarray}\label{sar2}
 \frac{Z_{\veQ-\veq_j-\veq_k}}{Z_{\veQ-\veq_j}}-\frac{Z_{\veQ-\veq_k}}{Z_{\veQ}}
&=& -\frac{\Lab^{b_k}\Las^{s_k}\Laq^{q_k}}{V} \left[ \frac{b_k b_j}{\Lab^2}M_{11} 
+\frac{s_k s_j}{\Las^2}M_{22}+\frac{q_k q_j}{\Laq^2} M_{33} \right. \nonumber\\
&& +\frac{b_k s_j + s_k b_j}{\Lab\Las} M_{12} + \frac{b_k q_j+ q_k b_j}{\Lab\Laq}
M_{13} \left. +\frac{s_k q_j+ q_k s_j}{\Las\Laq} M_{23} \right] + 
\mathcal{O}(V^{-2}) \nonumber \\
&\equiv& \frac{\lambda_k}{V} \, C_{jk} + \mathcal{O}(V^{-2}).
\end{eqnarray}
It is not difficult to realize, from their definition, that the matrix ${\sf M}$ 
and the $C_{jk}$ factors do not depend on the volume, i.e. they stay finite and 
non-vanishing in the thermodynamic limit. Therefore, we can recast 
Eq.~(\ref{MBomega}) as:
\begin{equation}\label{canom}
  \omega_h = 1 + \frac{\sum_{j\in h} \langle N_j \ranGC 
  \sum_{k\in h} \langle N_k \ranGC C_{jk}}
{V \sum_{j\in h} \langle N_j \ranGC} + \mathcal{O}(V^{-1}).
\end{equation}
where the second term on the right hand side of the above equation is finite in 
the limit $V \to \infty$, thus giving rise to non trivial values of the scaled variance 
$\omega$. 

In order to obtain a more intelligible expression of the extra-Poisson term in 
Eq.~(\ref{canom}), let us consider the scaled variance $\omega_j$ of a single hadron 
species in a simplified case, by neglecting baryon number and strangeness, i.e. 
considering only electric charge. The hessian matrix is trivially diagonal and one 
is left with only one element in the matrix ${\sf M}$, i.e.:
$$M_{33} = \left(\frac{\partial^2 f}{\partial w_{Q}^2} \Bigg|_{\vew_0}\right)^{-1} \; .$$
Then, neglecting possible hadrons with two or more units of electric charge, we
obtain the following expression for $C_{jk}$:
\begin{equation}\label{cjk}
 C_{jk} = \frac{q_k q_j}{\Laq^2} M_{33} =  \frac{q_k q_j}{\Laq^2} H_{33}^{-1}
  = \frac{q_k q_j}{\Laq^2}\frac{V \Laq^2}{Q + \sum_j \langle N_j \rangle_\GC 
  q_j (q_j - 1)} = \frac{V q_k q_j}{Q + 2 \langle h^- \rangle_\GC} = 
  \frac{V q_k q_j}{\langle h^+ \rangle_\GC + \langle h^- \rangle_\GC}
\end{equation}
where $\langle h^+ \ranGC$ and $\langle h^- \ranGC$ are the mean multiplicities of 
positive and negative hadrons respectively.  
By plugging the last expression into Eq.~(\ref{canom}) and taking into account that
for a single hadron species $j=k$ and $q_j^2 = 1$, we then obtain:
\begin{equation}\label{single}
 \lim_{V \to \infty} \omega_j = 1 + \frac{\langle N_j \ranGC}
 {\langle h^+ \rangle_\GC + \langle h^- \rangle_\GC}
\end{equation}
Likewise, it can be proved that, for the set of all negative, positive and charged
hadrons, scaled variances read:
\begin{eqnarray} \label{all}
\lim_{V \to \infty}\omega_\pm &\sim& 1-\frac{\langle h^\pm \ranGC}
{\langle h^+ \ranGC + \langle h^- \ranGC} \nonumber \\
\lim_{V \to \infty}\omega_\ch &\sim& 1-\left( 
\frac{\langle h^+ \ranGC - \langle h^- \ranGC}
{\langle h^+ \ranGC + \langle h^- \ranGC}\right)^2.
\end{eqnarray}
similarly to the classical pion gas case.
The Eq.~(\ref{single}) shows that the deviation from the Poisson statistics is 
proportional to the relative weight, in terms of multiplicity, of the species $j$ 
with respect to all species carrying a non-vanishing value of the same charge. 
This nicely meets our physical 
intuition that in a system with many species, the distribution of single and rarely 
produced ones, is little affected by global conservation laws. Conversely, according
to Eq.~(\ref{all}), the scaled variance of the most inclusive sets are strongly 
affected by the conservation laws and can show large deviations from Poisson statistics.
These general features hold in the more general case of three conserved charges,
although the simple formulae (\ref{single}) and (\ref{all}) have to be modified. 
Nevertheless, they can be taken as ``rules of thumb" to make estimates of scaled 
variances to a first approximation.

\subsection{Quantum statistics} 
\label{subs:quantum}

The generalization of previous expression to the case of quantum statistics is 
rather straightforward. The GC partition function reads:
\begin{equation}
  Z_{\GC} = 
  \prod_j \exp \left[ \sum_{n_j=1}^\infty \frac{\zjn \lambda_j^{n_j}}{n_j} \right]
\end{equation}
where:
\begin{equation}\label{zjn}
 \zjn = (\mp 1)^{n_j+1} \frac{(2J_j+1) V}{2\pi^2 n_j} T m_j^2 
  {\rm K}_2 \left(\frac{n_j m_j}{T}\right)
\end{equation}
with the upper sign for fermions and the lower for bosons. This yields the Hessian:
\begin{equation}
 H_{Q_l Q_n}(\vew_0) = \frac{1}{V \lambda_{Q_l}\lambda_{Q_n}} 
 \left[ Q_l\delta_{ln} + \sum_j \sum_{n_j=1}^\infty 
 q_{l,j} \left( n_j q_{n,j}-\delta_{kl} \right) \zjn \lambda_j^{n_j} \right\}.
\end{equation}
The first and second moments become:
\begin{eqnarray}\label{qmom} 
 \langle N_h \rangle &=& \sum_{j\in h} \langle N_j \rangle =  \sum_{j\in h} 
 \sum_{n_j=1}^\infty  \zjn \frac{Z_{\veQ-n_j\veq_j}}{Z_{\veQ}} \nonumber \\
\langle N^2_h \rangle &=& \sum_{j\in h} \sum_{n_j=1}^\infty n_j \zjn 
 \frac{Z_{\veQ-n_j\veq_j}}{Z_{\vec{Q}}} + \sum_{j\in h} \sum_{n_j=1}^\infty 
  \zjn \sum_{k \in h} \sum_{n_k=1}^\infty \zkn 
  \frac{Z_{\veQ-n_j\veq_j-n_k\veq_k}} {Z_{\vec{Q}}}.
\end{eqnarray}
Therefore:
\begin{equation}\label{qomega}
\omega_h = 1 + \frac{\displaystyle \sum_{j\in h} \sum_{n_j=1}^\infty (n_j-1) 
 \zjn \frac{Z_{\veQ-n_j\veq_j}}{Z_{\veQ}}}{\sum_{j \in h} \langle N_j \rangle} +
 \frac{\displaystyle \sum_{j \in h} \sum_{n_j=1}^\infty \zjn 
 \frac{Z_{\veQ-n_j\veq_j}}{Z_{\veQ}} \sum_{ k \in h} \sum_{n_k=1}^\infty \zkn
 \left(\displaystyle{\frac{Z_{\veQ-n_k\veq_k-n_j\veq_j}}{Z_{\veQ-n_j\veq_j}} - 
 \frac{Z_{\veQ-n_k\veq_k}}{Z_{\veQ}}}\right)}{\sum_{j \in h} \langle N_j \rangle}
\end{equation}
The middle term on the right hand side is the quantum correction to the Boltzmann
statistics and is kept in GCE, whilst the rightmost term represents the CE 
correction to the GC expression. As far as this last term is concerned, it can
be easily realized that the quantum hadron gas is equivalent to a classical 
hadron gas with an infinite number of particle species formed out of the actual
ones by multiplying their mass and charges by a positive integer number. Thus,
the same chain of arguments for the Boltzmann case can be followed in order to
derive an asymptotic expression of $\omega$:
\begin{equation}\label{qomega-asy}
\omega_h = 1 + \frac{\displaystyle \sum_{j\in h} \sum_{n_j=1}^\infty (n_j-1) 
 \zjn \lambda_j^{n_j}}{\sum_{j\in h} \langle N_j \rangle_\GC}
 + \frac{\displaystyle \sum_{j\in h} \sum_{n_j=1}^\infty n_j \zjn \lambda_j^{n_j}
  \sum_{k\in h} \sum_{n_k=1}^\infty n_k \zkn \lambda_k^{n_k} C_{jk}}
 {V \sum_{j\in h} \langle N_j \ranGC}+ \mathcal{O}(V^{-1}),
\end{equation}
where $C_{jk}$ has the same expression as in Eq.~(\ref{sar2}).

\section{Numerical results and Monte-Carlo calculations in the canonical ensemble}

In the previous section we have calculated the scaled variances of multiplicity
distributions in the CE in the thermodynamic limit by using saddle point 
asymptotic expansion. We can now provide numerical values of these variances 
for several cases of interest for ultra-relativistic heavy ion collisions. 
In our calculation, all light-flavoured hadron species up to a mass $\simeq 1.8$ GeV
quoted in the 2002 issue of Particle Data Book \cite{pdg2002} are included.
The needed intensive input parameters for these calculations are the temperature 
$T$ and the charge densities. The baryon density $\rho_B$ is varied between 0 and 
$0.3$ fm$^{-3}$, while the strangeness density $\rho_S$ is set to zero and the 
electric charge density is set to $\rho_Q = 0.4 \rho_B$, corresponding to the 
ratio $Z/A$ of Pb-Pb and Au-Au collisions. The chemical potentials and fugacities 
are determined accordingly.   

The asymptotic scaled variances of the multiplicity distributions of primary baryons, 
strange and negative particles, as well as primary $\pi^-$, p and K$^+$ for a 
completely neutral hadron gas are quoted in table \ref{cantab1}. The values 
in the CE in the Boltzmann approximation (B) are compared to the full quantum 
statistics ones in the CE (Q) and in the GCE (Q GC).
\begin{table}[htb]
\begin{center}
\caption{Scaled variances for the full ideal hadron-resonance gas in the CE 
for $T=120,160,180$ MeV and vanishing charge densities in the thermodynamic limit
for different sets of primary hadrons (baryons, strange particles, negatives, all, 
pions, protons and kaons). In the statistics row, ``B" stands for Boltzmann, ``Q"
for Quantum and ``Q GC" for quantum grand-canonical.} 
\label{cantab1}
\vspace{0.2cm}
\begin{tabular}{||c||c|c|c||c|c|c||c|c|c||}
\hline
$T$ [MeV] &  \multicolumn{3}{|c||}{120}  &
\multicolumn{3}{|c||}{160} & \multicolumn{3}{|c||}{180} \\
\hline statistics & B  & Q  & Q GC & B & Q & Q GC & B & Q & Q GC \\
\hline \hline $\omega_B$ &0.500&0.500&1.000&0.500&0.500&1.000
&0.500&0.500&1.000 \\
\hline $\omega_S$ &0.504&0.506&1.004&0.517&0.520&1.006
&0.524&0.528&1.007 \\
\hline $\omega_-$ &0.502&0.535&1.066&0.509&0.536&1.055
&0.512&0.535&1.045 \\
\hline $\omega_{\mathrm{tot}}$ &1&1.061&1.061&1&1.047&1.047
&1&1.038&1.038 \\
\hline $\omega_{\pi^-}$ &0.603&0.642&1.088&0.762&0.824&1.119
&0.831&0.911&1.131 \\
\hline $\omega_{\mathrm{p}}$ &0.892&0.892&1.000&0.931&0.931&0.999
&0.942& 0.941&0.998 \\
\hline $\omega_{K^+}$ &0.795&0.804&1.005&0.865&0.879&1.013
&0.896& 0.913&1.018 \\
\hline
\end{tabular}
\end{center}
\end{table}
It can be seen that $\omega_B$, $\omega_S$ and $\omega_-$ are essentially 1/2 
independently of the temperature, while $\omega_{\rm tot}=1$. This is not accidental, 
but a consequence of the vanishing of charge densities. To prove this, let us
consider the scaled variance $\omega_{B+\bar B}$ of baryons and antibaryons. 
For this special case, it can be realized that the numerator on the right hand 
side of Eq.~(\ref{canom}), i.e. the extra-Poisson contribution to the scaled 
variance, vanishes. The reason is simply that the factors $C_{jk}$ (see Eq.~(\ref{sar2}))
enter in the sum with opposite values for a baryon-antibaryon pair with 
respect to a baryon-baryon or antibaryon-antibaryon pair, while $\langle N_j \ranGC
\langle N_k \ranGC$ has the same value for all above combinations if the 
gas is neutral. Since there is an equal number of baryon-antibaryon and 
baryon-baryon or antibaryon-antibaryon pairs, the sum in the numerator of 
Eq.~(\ref{canom}) vanishes, thus $\omega_{B+\bar B} = 1$. In terms of the
variance of the random variables $N_B+N_{\bar B}$, we then have:
\begin{equation}
 \langle \delta (N_B + N_{\bar B})^2 \rangle = 
 \langle N_B \rangle + \langle N_{\bar B} \rangle = 2 \langle N_B \rangle 
\end{equation}
where we have used the neutrality condition. However, this condition holds for
the random variables themselves and not just for their mean values, that is 
$N_B = N_{\bar B}$. Hence, the left hand side can be written as 
$4 \langle \delta N_B^2 \rangle$, thence:
\begin{equation}
 \langle \delta (N_B + N_{\bar B})^2 \rangle = 
 4 \langle \delta N_B^2 \rangle = 2 \langle N_B \rangle  \;\; \Longrightarrow
 \;\; \omega_B = \frac{\langle \delta N_B^2 \rangle}{\langle N_B \rangle} = 
  \frac{1}{2}
\end{equation}
i.e. the scaled variance of baryons is 1/2. The same argument could be carried
over to $\omega_S$ and $\omega_{\pm}$ if there were only singly strange and charged 
hadrons. In fact, the slight difference from 1/2 for $\omega_S$ seen in 
table~\ref{cantab1} is due to multi-strange particles, while quantum statistics 
effects in primary pion production are the main responsible for the similar 
deviation of $\omega_-$.

For primary protons, pions and kaons, the scaled variance is not too far from
its Poissonian limit; this is a consequence of the overall large number of species 
in the hadron gas, in accordance with the argument at the end of 
Subsect.~\ref{subs:general}. As pion and kaons and protons are the lightest 
charged, strange and baryon particles respectively, they are relatively more 
abundantly produced at low temperatures than at high temperatures and this 
is reflected in the growth of $\omega$ towards its Poisson limit going from
$T=120$ MeV to $T=180$ MeV. 
\begin{table}[h]
\begin{center}
\caption{Scaled variances for the full ideal hadron-resonance gas in the CE 
with quantum statistics for different sets of primary hadrons (baryons, antibaryons,
strange, antistrange, positives, negatives, charged), in the thermodynamics limit,
as a function of $T$ and baryon density $\rho_B$. Strangeness density $\rho_S=0$ 
and electric charge density $\rho_Q = 0.4 \rho_B$.} 
\label{cantab2}
\vspace{0.2cm}
\begin{tabular}{||c||c|c|c||c|c|c||c|c|c||}
\hline
$T$ [MeV] &  \multicolumn{3}{|c||}{120}  &
\multicolumn{3}{|c||}{160} & \multicolumn{3}{|c||}{180} \\
\hline $\rho_B$ [fm$^{-3}$] &0.0&0.15&0.30&0.0&0.15&0.30&0.0&0.15&0.30  \\
\hline \hline $\omega_B$ & 0.500 & 0.005 & 0.008 &0.500 & 0.043
&0.015&0.500&0.195&0.085\\
\hline $\omega_{\bar{B}}$ &0.500 & 1.000 & 1.000 & 0.500&0.952
&0.980&0.500&0.799&0.905\\
\hline $\omega_S$ & 0.506 & 0.474 & 0.481&0.520&0.474
&0.465&0.528&0.507&0.485\\
\hline $\omega_{\bar{S}}$ & 0.506 &0.493 & 0.494&0.520&0.493
&0.486&0.528&0.514&0.498\\
\hline $\omega_+$ & 0.535 &0.255& 0.209 &0.536&0.391
&0.316&0.535&0.464&0.390 \\
\hline $\omega_-$ & 0.535 & 0.571 & 0.566 &0.536&0.555
&0.549&0.535& 0.562&0.558\\
\hline $\omega_\ch$ & 1.066 &0.649 & 0.543 &1.055&0.872
&0.749&1.045&0.974&0.872\\
\hline
\end{tabular}
\end{center}
\end{table}

In table \ref{cantab2} we quote the asymptotic scaled variances of 
particles and anti-particles for different temperatures and baryon densities,
for full quantum statistics.
It can be seen that $\omega_B$ and $\omega_{\bar{B}}$ are found to quickly 
converge to the limits 0 and 1 respectively with increasing $\rho_B$, as predicted 
by the rule of thumb for the baryons analogous to Eq.~(\ref{all}). On the other 
hand, the strange sector is dominated by the strangeness neutrality condition and 
is weakly dependent on baryon density. The fluctuations of positive hadrons are 
dominated by protons at high baryon fugacities. This is reflected in a monotonically 
decreasing value $\omega_+$ with increasing $\rho_B$. Conversely, on the basis of
the rule of thumb~(\ref{all}), one would expect an equally strong increase of 
$\omega_-$. Surprisingly, this is not the case because at high $\lambda_B$ the 
$\Delta^-$ becomes the dominant negative hadron and that simple approximation 
fails because of the strong correlation between electric charge and baryon number.  

\subsection{Monte-Carlo simulations}

The scaled variances determined by means of analytical calculations have been 
compared with those obtained through Monte-Carlo simulations. The basic idea of 
this method is to extract randomly $K$-uples ${\Nj}$ of multiplicities $N_j$ for 
each hadron species $j$ according to the multi-species multiplicity distribution of 
the canonical ensemble and averaging therafter. This method allows to determine 
numerically, with a finite statistical 
error, not only scaled variances but also higher order moments and, in general, to 
visualize the shape of the distributions. Furthermore, this method makes it possible 
to make calculations at final hadron level, taking into account resonance decays and 
thereby allowing a comparison of theoretical calculations with actual measurements.  
The multi-species multiplicity distribution in the canonical ensemble has been 
determined in the form of a cluster decomposition in ref.~\cite{becaferro2}: 
\begin{equation}\label{candis}
 P({\Nj}) = \frac{1}{Z_{\veQ}} \left[ \prod_j \sum_{\hpartj} 
 \prod_{n_j=1}^{N_j} \frac{z_{j(n_j)}^{h_{n_j}}}{n_j^{h_{n_j}} h_{n_j}!} \right] 
 \delta_{\veQ,\sum_j N_j \veq_j}
\end{equation}
where ${\hpartj}$ are partitions of the integers $N_j$ in the multiplicity representations, 
i.e. such that $N_j = \sum_{n_j=1}^{N_j} n_j h_{n_j}$; $H_j=\sum_{n_j=1}^{N_j} h_{n_j}$; 
and $z_{j(n_j)}$ are given in Eq.~(\ref{zjn}). 

In the limit of the Boltzmann statistics, the distribution (\ref{candis}) reduces
to a product of independent Poisson distributions, one for each species, with the 
constraint of charges conservation $\delta_{\veQ,\sum_j N_j \veq_j}$. A direct sampling of 
the distribution (\ref{candis}) is very difficult though. The most effective
method is the importance sampling technique, in which each event (namely 
a $K$-uple $\Nj$) is weighted by the ratio $w$ of the true distribution $P(\Nj)$ 
(\ref{candis}) and the actually sampled distribution $R(\Nj)$. The latter should be
a distribution quickly and efficiently sampled and, moreover, as similar as possible to 
$P(\Nj)$ to minimize statistical errors. In our case, we have chosen $R(\Nj)$ as the 
product of unconstrained Poisson distributions, 
like in Eq.~(\ref{poisson}). Their mean multiplicities are chosen to be those of 
the GCE, that is $\langle N_j \rangle = \zjl \lambda_j$, where the fugacities 
$\lambda_j$ are determined according to the saddle-point equations (\ref{saddle}); 
thereby, the mean values of the Poisson distributions in $R(\Nj)$ coincide with 
the actual CE average multiplicities in the thermodynamic limit.

The average of any observable $O$ after $N_S$ random samples reads:
\begin{equation}\label{average}
\langle O \rangle = \frac{\displaystyle \sum_{i=1}^{N_S} O_i w_i}
{\displaystyle \sum_{i=1}^{N_S} w_i} \qquad {\rm with} 
\;\; w_i = \frac{P(\Nj_i)}{R(\Nj_i)} \;\; ;
\end{equation}
statistical errors can be calculated accordingly \cite{becaferro2}. 

Unfortunately, with this method, it is not possible to calculate observables 
straight in the thermodynamic limit because the simulation can be carried out only
with a finite volume. Instead, one can study the variation of some observable of
interest as a function of the volume, fixing total charges or charge densities,
and estimate the thermodynamic limit by extrapolating. It should be pointed out 
that a too large volume cannot be used
in order not to diminish too much the efficiency of the Monte-Carlo calculation.
In fact, according to Eq.~(\ref{average}), only the $K$-uples with non-vanishing
weight, i.e. fulfilling $\veQ = \sum_j N_j \veq_j$ actually contribute to the 
average. On the other hand, random samples are extracted from $R(\Nj)$, which
is a product of Poisson distributions, and only a small fraction of them will
meet the charge conservation constraint. This fraction is the efficiency of the
sampling and reads:
\begin{equation}
 \eta = \sum_{\Nj} \prod_j \frac{1}{N_j!} \langle N_j \rangle_\GC^{N_j} 
 \e^{-\langle N_j \rangle_\GC} \delta_{\veQ,\sum_j N_j \veq_j}
\end{equation}
It can be easily proved, by developing the above expression, that the efficiency 
decreases as volume increases. In other words, the larger the volume, the more 
difficult is to catch a configuration which fulfills exactly the charge conservation
constraint. 
Then, to keep the statistical error small, one should not increase the volume too 
much. Fortunately, the statistical error on most averages decreases as volume 
increases because of the increase in multiplicity of single events, so that a lower
efficiency at larger volumes does not spoil the accuracy.

\begin{figure}[t]
\begin{center}
$\begin{array}{c@{\hspace{1in}}c}
    \epsfxsize=4.in
    \epsffile{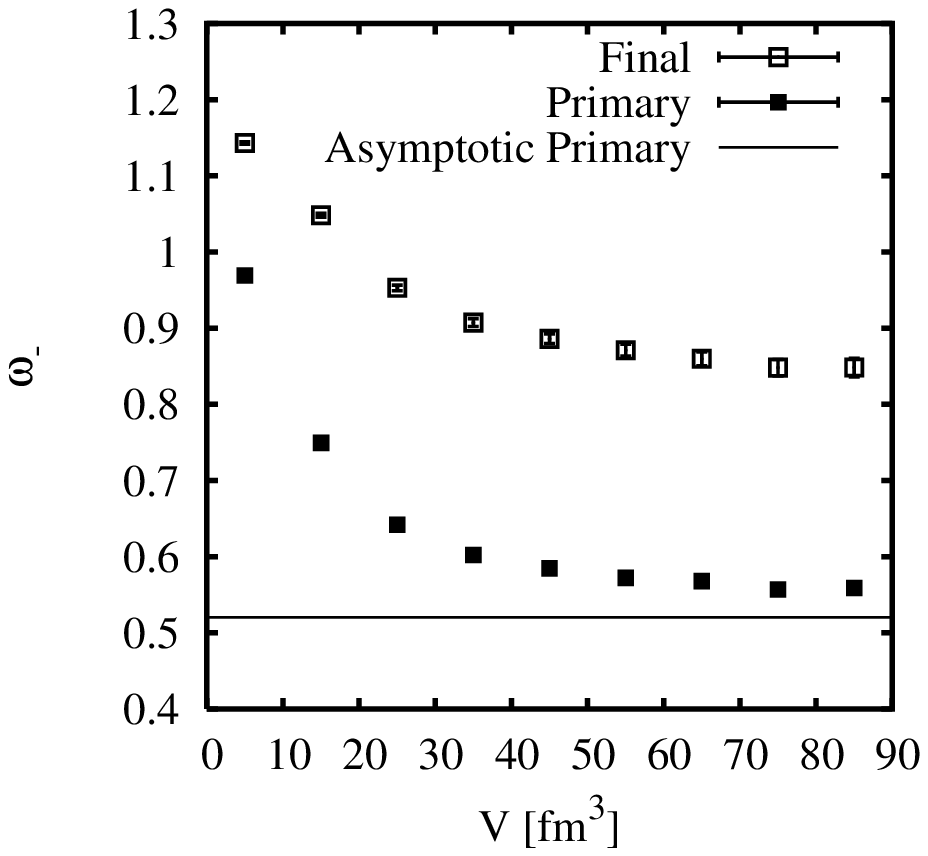} & \hspace*{-4.5cm}
        \epsfxsize=4.in
        \epsffile{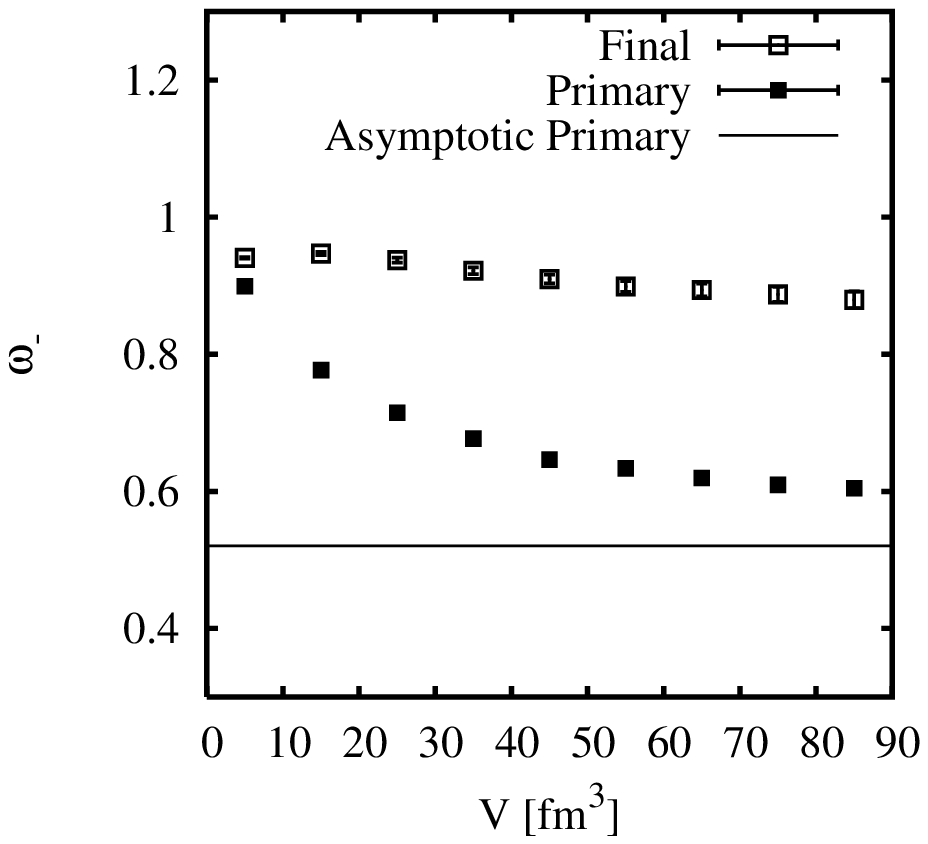} \\
\end{array}$
\vspace{-0.7cm}
\end{center}
\caption{Scaled variances in the canonical ensemble of the full ideal 
hadron-resonance gas for negative hadrons at $T=160$ MeV as functions of 
volume calculated with Monte-Carlo simulations. Closed squares are for 
primary particles and open squares are for final particles after strong 
and electromagnetic decays. The horizontal solid lines indicate the 
thermodynamic limit $V\rightarrow\infty$ calculated with asymptotic expansions.
Left panel: $e^+e^-$ charge configuration (Q=S=B=0). Right panel: pp charge 
configuration (Q=B=2, S=0). }
 \label{canfig1}
\end{figure}

\begin{figure}[h]
\begin{center}
$\begin{array}{c@{\hspace{1in}}c}
    \epsfxsize=4.in
    \epsffile{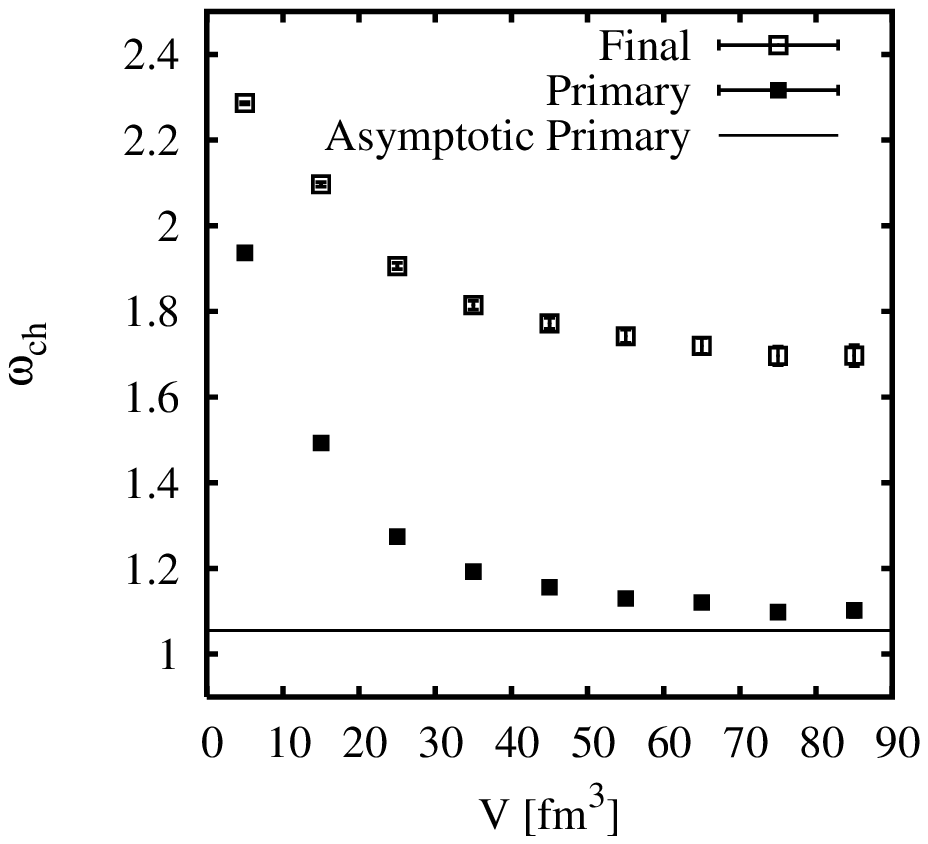} & \hspace*{-4.5cm}
        \epsfxsize=4.in
        \epsffile{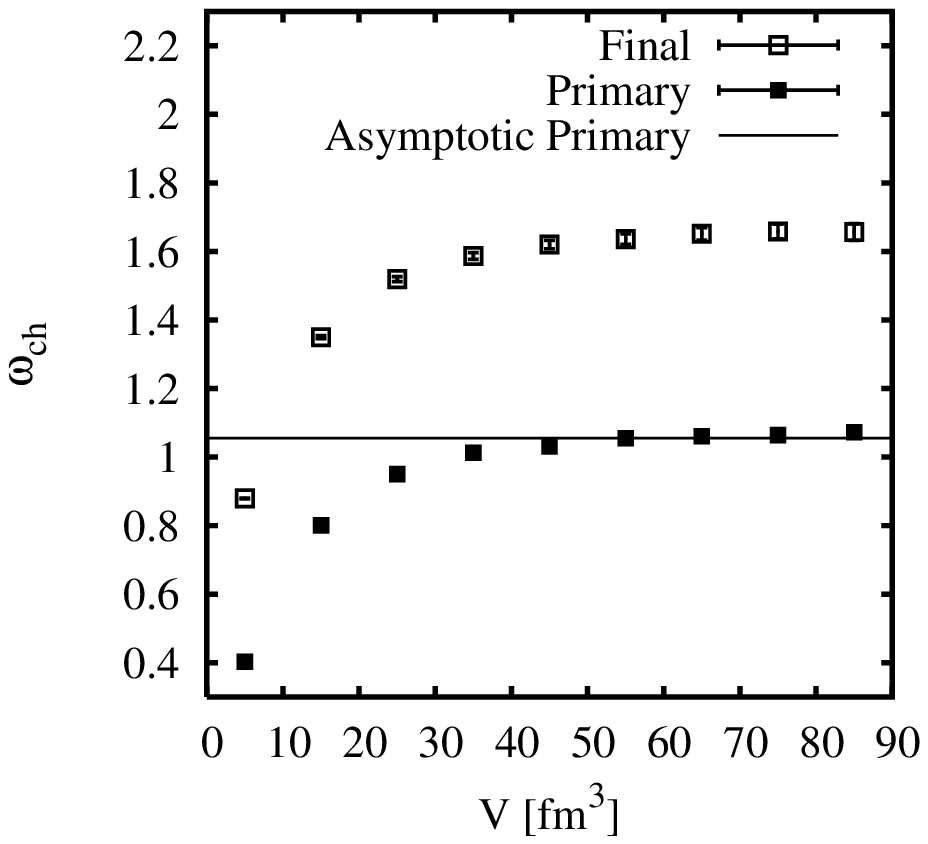} \\
\end{array}$
\vspace{-0.7cm}
\end{center}
\caption{Scaled variances in the canonical ensemble of the full ideal 
hadron-resonance gas for charged hadrons at $T=160$ MeV as functions of 
volume calculated with Monte-Carlo simulations. Closed squares are for 
primary particles and open squares are for final particles after strong 
and electromagnetic decays. The horizontal solid lines indicate the 
thermodynamic limit $V\rightarrow\infty$ calculated with asymptotic expansions.
Left panel: $e^+e^-$ charge configuration (Q=S=B=0). Right panel: pp charge 
configuration (Q=B=2, S=0).}
\label{canfig2}
\end{figure}

We have run simulations of canonical systems with vanishing charges (neutral systems)
and with charges equal to those of a pp collision, i.e. $Q=2,B=2,S=0$ (pp-like system)
for $T=160$ MeV and volumes $V$ ranging from 5 to 85 fm$^3$ in steps of 10 fm$^3$;
for each point, $10^7/V ({\rm fm}^3)$ effective samples (i.e. fulfilling charge 
conservation) have been drawn from $R(\Nj)$ with efficiencies varying from 26\% to 
0.2\% for neutral system and from 3\% to 0.2\% for pp-like system. In figs.~\ref{canfig1}, 
~\ref{canfig2} the obtained scaled variances are shown for negative and charged hadrons 
respectively, both at primary and final level. It can be seen that scaled variances 
slowly converge to their thermodynamic limits which are attained up to $\sim 5\%$ 
within $V \sim 100$ fm$^3$ and that those of the finals are always larger than 
primaries'. The latter effect is mainly owing to neutral particles decaying into a 
pair of charged particles. 

\begin{figure}[h]
\begin{center}
$\begin{array}{c@{\hspace{1in}}c}
    \epsfxsize=4.in
    \epsffile{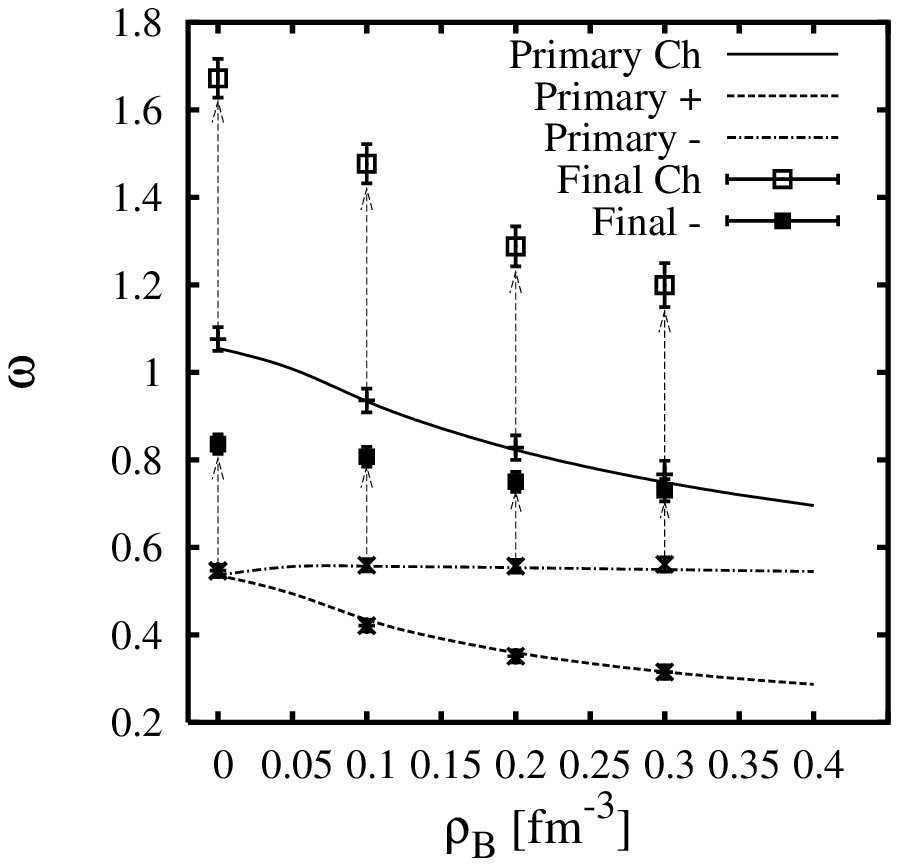} & \hspace*{-4.5cm}
        \epsfxsize=4.in
        \epsffile{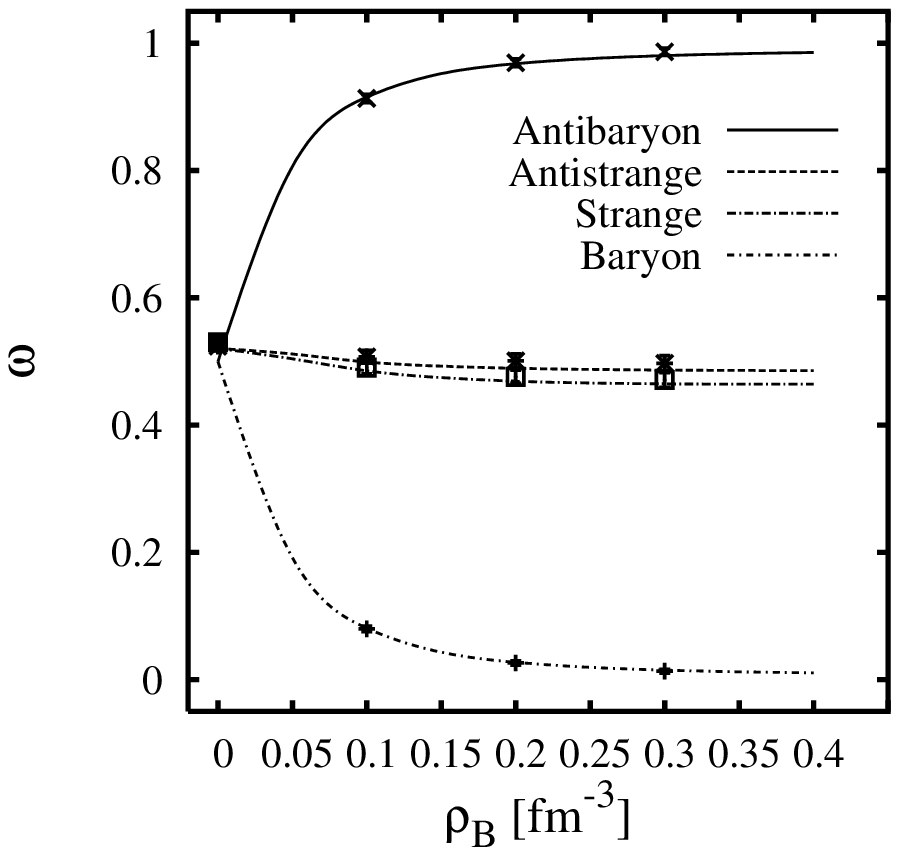} \\
\end{array}$
\vspace{-0.7cm}
\end{center}
\caption{Scaled variances in the canonical ensemble of the full ideal 
hadron-resonance gas for different sets of hadrons at $T=160$ MeV, $S=0$ 
and $Q/B=0.4$ as functions of baryon density $\rho_B$. Closed dots indicate
the calculated values with Monte-Carlo simulations at primary level, open
dots at final level with $V=200$ fm$^3$. The lines depict the thermodynamic 
limits indendently calculated with the asymptotic expansion formulae.
Left panel: charged, positive and negative hadrons; the arrows show the 
change in $\omega_-$ and $\omega_\ch$ from primary to final level.
Right panel: baryons, antibaryons, strange and antistrange hadrons.}
\label{canfig3}
\end{figure}

We have also calculated the scaled variances in the canonical ensemble for larger
systems and compared them with relevant analytical calculations. This is shown in 
fig.~(\ref{canfig3})
where we have plotted $\omega$ for different sets of particles as a function of
the baryon density $\rho_B$, with $\rho_Q=0.4 \rho_B$ and $\rho_S=0$. Monte-Carlo
results are shown as dots with errors bars and have been obtained by drawing $10^5$ 
effective samples with $V=200$ fm$^3$ for baryon densities 0., 0.1, 0.2, 0.3 fm$^{-3}$.
Due to the large volume, which is needed to ensure the effective reaching of the
thermodynamic limit, the efficiency of these runs is very low, in the range $(1-6)
\cdot 10^{-4}$. It can be clearly seen that Monte-Carlo and analytical calculations
are in excellent agreement.

\section{Asymptotic fluctuations in the microcanonical ensemble}

In principle, fluctuations in the microcanonical ensemble of the ideal hadron-resonance
gas in the thermodynamic limit can be studied by applying the same saddle-point 
expansion described in Sect.~2 for the canonical partition function to the 
microcanonical partition function (MPF), defined as \cite{becaferro1}:
\begin{equation}\label{micro}
\Omega = \sum_{h_V}\langle h_V | 
\delta^4(P-P_{\mathrm{op}})\delta_{\veQ,\veQ_{\mathrm{op}}}
| h_V \rangle.
\end{equation}
$| h_V \rangle$ being localized multi-hadronic states within the volume $V$. 
However, for the sake of simplicity, we will confine ourselves to energy conservation only, 
i.e. momentum conservation will be disregarded. Furthermore, we will discuss only
the case of Boltzmann statistics. Therefore, the operator $\delta^4(P-P_{\mathrm{op}})$ 
in Eq.~(\ref{micro}) is henceforth understood to be replaced with $\delta(E-E_{\mathrm{op}})$.
In this case, the partition function can be written as \cite{becaferro1}:
\begin{equation}\label{mpf}
\Omega_{\vec{Q}} = \lim_{\epsilon \to 0^+} \frac{1}{(2\pi\ii)^4}
 \int_{-\ii \infty+ \epsilon}^{+\ii \infty+\epsilon} 
  \!\!\!\! \dii w_E \oint \dii w_B \oint \dii w_S \oint \dii w_Q \; 
  w_E^{E-1} w_B^{-B-1} w_S^{-S-1} w_Q^{-Q-1} 
  \exp\!\left\{ V \sum_j F^{(0)}_j w_B^{b_j} w_S^{s_j} w_Q^{q_j}\right\},
\end{equation}
where the following convention is adopted:
\begin{equation}\label{ffunct} 
 F^{(n)}_j \equiv F^{(n)}_j(w_E) \equiv \frac{(2J_j+1)}{(2\pi)^3} \int \dii^3 p \; 
 \varepsilon_j^n w_E^{-\varepsilon_j}
\end{equation}
$\varepsilon_j$ being the energy of the $j^{\rm th}$ particle. The integral 
expression of the microcanonical partition function (\ref{mpf}) looks very similar 
to the canonical partition function (\ref{canpart}), with a further integration 
over the imaginary line $w_E=\ii \epsilon$. Insertion of the fictitious fugacities 
$\lambda_j$ allow to calculate the moments like in the CE in Eq.~(\ref{moments}). 
Denoting by:
\begin{equation}\label{omegaj}
\Omega^{(j_1,\ldots,j_K)}_{\veQ} \equiv \frac{1}{(2\pi\ii)^4} 
 \int \left[\prod_i \dii w_i \right] F^{(0)}_{j_1} \ldots F^{(0)}_{j_K} 
  w_E^{E-1} w_B^{-B-1} w_S^{-S-1} w_Q^{-Q-1} \exp\!\left\{V \sum_k F^{(0)}_k 
  w_B^{b_k} w_S^{s_k} w_Q^{q_k}\right\}
\end{equation}
it is quite straightforward to derive the moments:
\begin{eqnarray} \label{mcmoments}
\langle N_h \rangle &=& V \sum_{j\in h} \frac{\Omj_{\veQ-\veq_{j}}}{\Omega_{\veQ}} \\
\langle N_h^2 \rangle &=& V \sum_{j\in h} \frac{\Omj_{\veQ-\veq_{j}}}{\Omega_{\veQ}}
+  V^2 \sum_{j,k\in h} \frac{\Omjk_{\veQ-\veq_{j}-\veq_{k}}}{\Omega_{\veQ}}
\end{eqnarray}
Consequently:
\begin{equation} \label{mcomega}
\omega_h = 1 + V \frac{\displaystyle \sum_{j\in h} \frac{\Omj_{\veQ-\veq_j}}
{\Omega_{\vec{Q}}} \sum_{k\in h} \left( \frac{\Omjk_{\veQ-\veq_k-\veq_j}}
{\Omj_{\veQ-\veq_j}} - \frac{\Omk_{\veQ-\veq_k}}{\Omega_{\vec{Q}}} \right)}
{\displaystyle \sum_{j\in h} \frac{\Omj_{\veQ-\veq_j}}{\Omega_{\vec{Q}}}}.
\end{equation}
The whole procedure of saddle-point expansion for the canonical partition function
can be carried over here for the microcanonical partition function. The saddle point 
equation in the variable $w_E$ leads to the definition of the temperature $T=1/\beta$ 
\cite{becaferro1} which is such that:
\begin{eqnarray}\label{saddle2}
 && \sum_j \frac{F^{(1)}_j(\e^\beta)}{V} \Lab^{b_j}\Las^{s_j}\Laq^{q_j}
  = \rho_E \nonumber \\
 && \sum_j q_{l,j} \frac{F^{(0)}_j(\e^\beta)}{V} \Lab^{b_j}\Las^{s_j}\Laq^{q_j}
 = \frac{Q_l}{V} \qquad l=1,2,3 
\end{eqnarray}  
The appearance of an exponential of a dimensional quantity in the equations
will not cause any problem because $\e^\beta$ will always have energies as
exponents in the final expressions. The mean multiplicity (\ref{mcmoments}) converges 
to the corresponding grand-canonical value in the thermodynamic limit (see Appendix B).
Conversely, the term within brackets in the scaled variance expression (\ref{mcomega})
is ${\mathcal O}(V^{-1})$ yielding a non-trivial value in the thermodynamic limit. 
In fact, the following expression is obtained for $\omega$ for $V \to \infty$:
\begin{eqnarray}\label{mctl}
\lim_{V \to \infty} \omega_h = 1 &-& \frac{1}{\sum_{j\in h} \langle N_j \ranGC}
\sum_{j\in h} \langle N_j \ranGC \sum_{k\in h} \lambda_k 
\left\{ \frac{\hF_j^{(1)}\hF_k^{(1)}}{\hF_j^{(0)}}\e^{-2\beta} M_{00}\right. 
\nonumber\\
&& -\sum_{l=1}^3 \left[q_{l,k} \hF_j^{(1)}\hF_k^{(0)}+q_{l,j} \hF_j^{(0)}\hF_k^{(1)} 
\right] \frac{\e^{-\beta} \lambda_{Q_l}^{-1}} {\hF_j^{(0)}} M_{0l} 
\left. + \sum_{l,n=1}^3 q_{l,j}q_{m,k} \hF_k^{(0)}\lambda_{Q_l}^{-1}
 \lambda_{Q_n}^{-1} M_{ln}\right\}.
\end{eqnarray}
where $\hF^{(n)}\equiv F^{(n)}(\e^{\beta})$ and, as usual, 
$M_{ln} = \sum_{i=0}^3 A_{il} A_{in}/h_i$, the index $0$ pertaining to the 
energy-related variable $w_E$.
The details of the derivation of Eq.~(\ref{mctl}) can be found in Appendix B.

\subsection{Numerical calculations}

The formula (\ref{mctl}) has been used to evaluate the scaled variance in the MCE
in the thermodynamic limit. In table~\ref{microtab1} we compare the scaled variance
of several sets of hadrons with the corresponding value in the CE, for different 
temperatures. It is seen that that $\omega$ is further reduced owing to the additional 
constraint of energy conservation. This reduction is more considerable for larger 
sets of hadrons and attains its maximum for all primary hadrons ($\omega_{tot}$). 

\begin{table}[htb]
\begin{center}
\caption{Scaled variances for the full ideal hadron-resonance gas in the MCE 
and CE for $T=120,160,180$ MeV and vanishing charge densities in the thermodynamic limit
for different sets of primary hadrons (baryons, strange particles, negatives, all, 
pions, protons and kaons). Momentum conservation and 
quantum statistics have been disregarded.}
\label{microtab1}
\vspace{0.2cm}
\begin{tabular}{||c||c|c||c|c||c|c||}
\hline
$T$ [MeV] &  \multicolumn{2}{|c||}{120}  &
\multicolumn{2}{|c||}{160} & \multicolumn{2}{|c||}{180} \\
\hline ensemble &MCE&CE&MCE&CE&MCE&CE \\
\hline \hline $\omega_B$ &0.426&0.500&0.340&0.500&0.311&0.500 \\
\hline $\omega_S$ &0.357&0.504&0.349&0.517&0.346&0.524 \\
\hline $\omega_-$ &0.307&0.502&0.310&0.509&0.304&0.512 \\
\hline $\omega_\ch$ &0.610 & 1 & 0.603 &1&0.584&1 \\
\hline $\omega_{\mathrm{tot}}$ &0.257&1&0.207&1&0.172&1 \\
\hline $\omega_{\pi^-}$ &0.523&0.603&0.736&0.762&0.815&0.831 \\
\hline $\omega_{\mathrm{p}}$&0.882 &0.892&0.920&0.931&0.932&0.942 \\
\hline $\omega_{K^+}$&0.756 &0.795&0.847&0.865&0.883&0.896 \\
\hline
\end{tabular}
\end{center}
\end{table}

\begin{table}[htb]
\begin{center}
\caption{Scaled variances for the full ideal hadron-resonance gas in the MCE 
for different sets of primary hadrons (baryons, antibaryons, strange, antistrange, 
positives, negatives, charged), in the thermodynamics limit,
as a function of $T$ and baryon density $\rho_B$. Strangeness density $\rho_S=0$ 
and electric charge density $\rho_Q = 0.4 \rho_B$. Momentum conservation and quantum 
statistics have been disregarded.} 
\label{microtab2}
\vspace{0.2cm}
\begin{tabular}{||c||c|c|c||c|c|c||c|c|c||}
\hline
$T$ [MeV] &  \multicolumn{3}{|c||}{120}  &
\multicolumn{3}{|c||}{160} & \multicolumn{3}{|c||}{180} \\
\hline $\rho_B$ [fm$^{-3}$] &0.0&0.15&0.30&0.0&0.15&0.30&0.0&0.15&0.30  \\
\hline \hline $\omega_B$& 0.426 & 0.000 &  0.000 & 
0.340 & 0.036& 0.012&
0.311 &0.133&  0.064\\
\hline $\omega_{\bar{B}}$& 0.426 & 0.996 & 0.998&
0.340 & 0.808&  0.896&
0.311 &0.543&  0.686\\
\hline $\omega_S$ & 0.357 & 0.356 & 0.362&
0.349 & 0.341 & 0.343&
0.346 & 0.341 & 0.340 \\
\hline $\omega_{\bar{S}}$& 0.357 & 0.367& 0.369&
 0.349 &  0.360 & 0.362&
0.346 & 0.354 & 0.358 \\
\hline $\omega_+$ & 0.307& 0.192 & 0.172&
 0.310& 0.246 & 0.218&
0.304 & 0.267 & 0.242 \\
\hline $\omega_-$& 0.307 & 0.438&  0.464&
0.310 & 0.374 & 0.403&
0.304 & 0.342& 0.369 \\
\hline $\omega_\ch$& 0.610 & 0.484& 0.440&
0.603 & 0.562 & 0.524&
0.584 & 0.572 & 0.551 \\
\hline $\omega_{\mathrm{tot}}$& 0.257 & 0.141 & 0.107&
0.207 & 0.173 & 0.148&
0.172 & 0.163& 0.150\\
\hline
\end{tabular}
\end{center}
\end{table}

\begin{figure}[h]
\begin{center}
$\begin{array}{c@{\hspace{1in}}c}
    \epsfxsize=4.in
    \epsffile{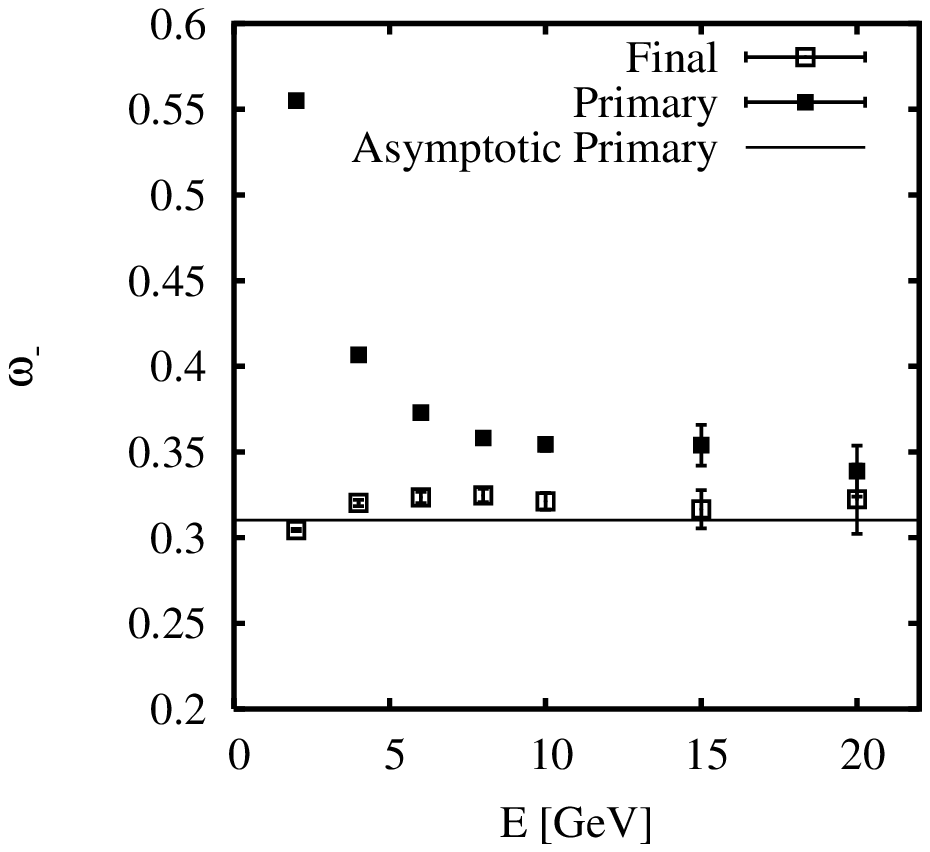} & \hspace*{-4.5cm}
        \epsfxsize=4.in
        \epsffile{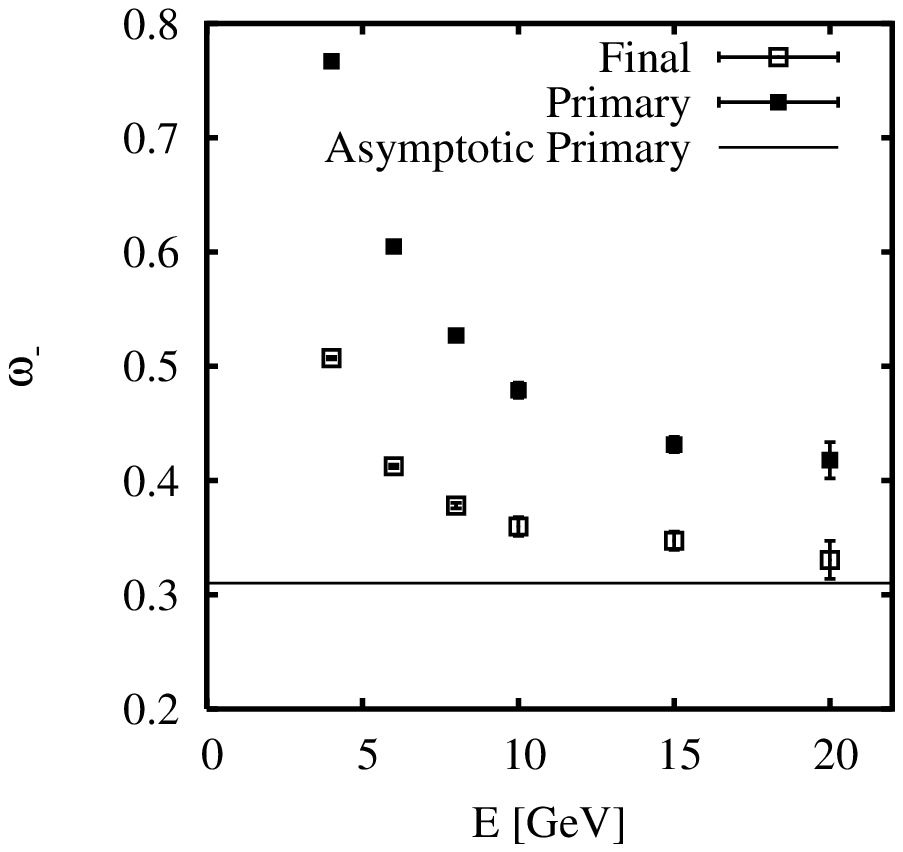} \\
\end{array}$
\vspace{-0.7cm}
\end{center}
\caption{Scaled variances in the microcanonical ensemble of the full ideal 
hadron-resonance gas for negative hadrons at an energy density of 0.4 GeV/fm$^3$ 
as functions of energy calculated with Monte-Carlo simulations. Closed squares 
are for primary particles and open squares are for final particles after strong 
and electromagnetic decays. The horizontal solid lines indicate the corresponding
thermodynamic limit $V\rightarrow\infty$ calculated with asymptotic expansions,
the resulting temperature being $\simeq 160$ MeV.
Left panel: $e^+e^-$ charge configuration (Q=S=B=0). Right panel: pp charge 
configuration (Q=B=2, S=0).}
\label{microfig1}
\end{figure}

\begin{figure}[h]
\begin{center}
$\begin{array}{c@{\hspace{1in}}c}
    \epsfxsize=4.in
    \epsffile{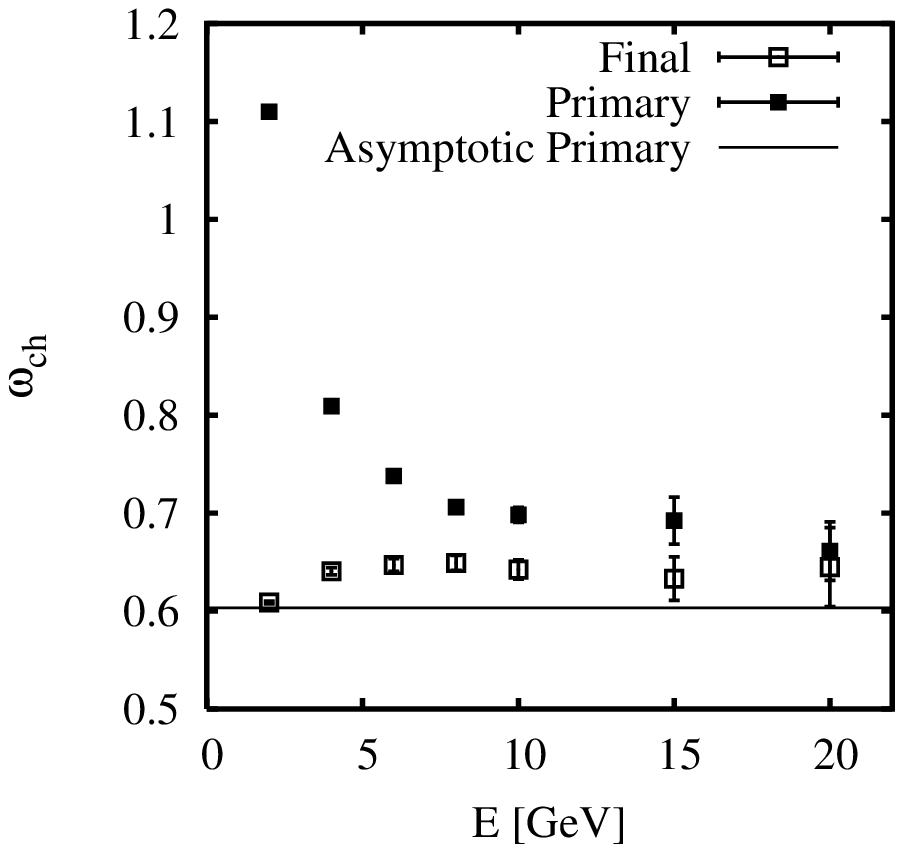} & \hspace*{-4.5cm}
        \epsfxsize=4.in
        \epsffile{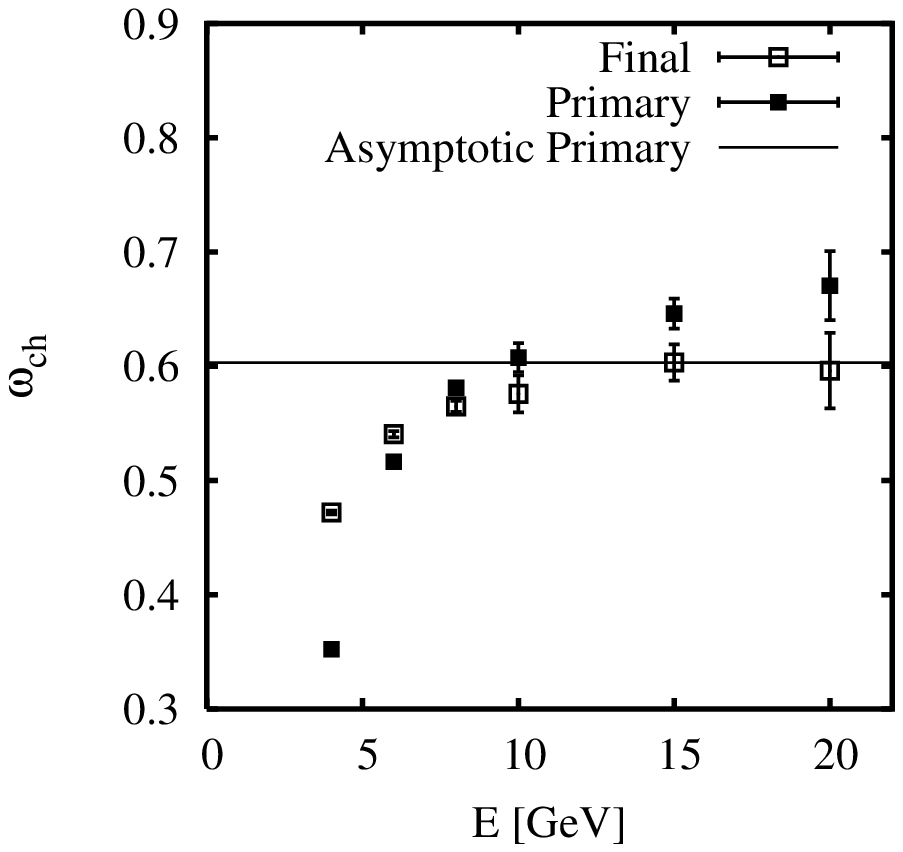} \\
\end{array}$
\vspace{-0.7cm}
\end{center}
\caption{Scaled variances in the microcanonical ensemble of the full ideal 
hadron-resonance gas for negative hadrons at an energy density of 0.4 GeV/fm$^3$ 
as functions of energy calculated with Monte-Carlo simulations. Closed squares 
are for primary particles and open squares are for final particles after strong 
and electromagnetic decays. The horizontal solid lines indicate the corresponding
thermodynamic limit $V\rightarrow\infty$ calculated with asymptotic expansions,
the resulting temperature being $\simeq 160$ MeV.
Left panel: $e^+e^-$ charge configuration (Q=S=B=0). Right panel: pp charge 
configuration (Q=B=2, S=0).}
\label{microfig2}
\end{figure}

Also in the MCE, there are available Monte-Carlo algorithms to make
numerical calculations of particle distributions both at primary and final levels.
We have used an optimized importance sampling algorithm described in detail in 
ref.~\cite{becaferro2}, which is the state-of-art in this field.
Unlike in the case of CE, Monte-Carlo simulations in the MCE of the full ideal 
hadron-resonance gas cannot be performed beyond values of $V \approx 50$ fm$^3$ and 
energies larger than $\approx 20$ GeV, with present computing power and numerical 
techniques \cite{becaferro2}. These values are not large enough to make an accurate
numerical assessment of the thermodynamic limit. This can be seen from 
figs.~\ref{microfig1},~\ref{microfig2}, where the Monte-Carlo calculated scaled 
variances for increasing energies have been compared to their thermodynamic limit.
Although at $E=20$ GeV central values sizeably differ from the thermodynamic limit, 
yet the overall trend is quite clear. It should be noted that, unlike in the asymptotic 
expansions, in the Monte-Carlo also momentum conservation was enforced, so that 
there might be a residual effect due to this difference in the calculation scheme.

From a closer look at figs.~\ref{microfig1},~\ref{microfig2}, it can be realized 
that, unlike in the CE case, the scaled variance of negatives and charged is larger
for primary than final hadrons especially at low energies and volumes. This might
seem surprising, as our first intuition is that the decays of primary neutral particles 
into charged introduce a further source of fluctuation, thus leading to an increase 
of $\omega$. This is indeed the case for the
CE. However, in the MCE there is a further constraint, namely that the minimum
number of primary particles is two. In a completely neutral MCE at very low energy
particles are essentially produced only in pairs because the channels with 3, 4 etc.
particles have negligible probability \cite{becaferro2}, thus one can have zero
(when two neutral particles are emitted) or one negative hadron (when one positive
and one negative particle are emitted) at primary level. It is not difficult to 
realize that the relative probability of these two events is 1/2 if there is an 
equally large number of positive, neutral and negative species and that the 
corresponding scaled variance is 2/3, what seems indeed to be the limit of $\omega_-$
at small energy value in fig.~\ref{microfig1}. Conversely, at final level, if 
all neutral particles decayed into one positive and one negative particle, there 
would always be one negative particle and the scaled variance would vanish. The
non-vanishing value is owing to neutral stable particles, neutral particle
decaying into stable neutral particles (e.g. $\pi^0 \rightarrow \gamma \gamma$) 
and negative particles giving rise to more than one negative particle in their
decays. Still, a strong reduction of scaled variance for final negatives is implied. 
A similar argument applies to charged particles.

\section{Discussion}

We have found that the thermodynamic limit of the scaled variance is different
in different ensembles. This effect has been understood for a long time in statistical
mechanics. Though, from the previous derivations, the reader might have had the 
impression that this inequivalence between GCE, CE and MCE is a long-reaching 
consequence of a complicated analytical work not driven by a clear physical insight.
Recently, it has been pointed out \cite{turko} that variances are qualitatively 
different from particle multiplicities in that their proportionality to the volume 
is not ``primordial'' but arises from the difference of two quantities whose leading
term is ${\cal O}(V^2)$ (see Eq.~(\ref{omega})). As a consequence, the behaviour
of variance in the thermodynamic limit is determined by sub-leading terms in both 
$\langle N^2 \rangle$ and $\langle N \rangle^2$ and different limits can be expected
in different ensembles. We would like to point out here that the different behaviour
in the thermodynamic limit can indeed be understood more simply and with more physical 
insight, by observing that, unlike particle multiplicities, variances are non-additive 
quantities in both the CE and MCE. 

The conceptual difference between extensivity and additivity has been recently 
discussed in ref. \cite{touch}. An additive quantity $X$ is such that, if we split 
a general system into $N$ subsystem:
\begin{equation}
  X = \sum_{i=1}^N X_i
\end{equation}
On the other hand, an extensive quantity is such that the limit:
\begin{equation}\label{intensive}
  \lim_{N \to \infty} \frac{X}{N} = x
\end{equation}
has a non-vanishing and finite value. If a quantity is additive, it is also extensive 
except for some exceptional case \cite{touch}. Conversely, extensivity does 
{\em not} imply additivity. Quantities which are extensive but not additive are 
defined as {\em pseudo-extensive} and their corresponding limit $x$ in Eq.~
(\ref{intensive}) {\em pseudo-intensive} \cite{touch}. 
It can be shown quite easily that additive quantities have the same thermodynamic 
limit in all ensembles. In fact, if we split a CE or MCE with a very large volume
into a large number of $N$ parts with volume $V/N$, each part is, by definition, 
a GCE with the rest of the system acting as a reservoir; this is just the way the
GCE is introduced in most statistical mechanics textbooks. Consequently, any $X_i$,
where $i$ labels a subsystem in the CE or MCE, 
has the same value as in the GCE with volume $V/N$ in the limit $V, N \to \infty$.
In other words, for any $i$: 
\begin{equation}\label{limgce}
  \lim_{V, N \to \infty} X_i =  \lim_{V, N \to \infty} \frac{X_{GCE}(V)}{N}
\end{equation}
If $X$ is additive, then:
\begin{equation}\label{tlim1}
  \lim_{V, N \to \infty} \sum_{i=1}^N X_i = \lim_{V \to \infty} X_{CE,MCE}(V)
\end{equation}
On the other hand, the left hand side of the previous expression also yields,
according to (\ref{limgce})
\begin{equation}\label{tlim2}  
  \lim_{V, N \to \infty} 
   \sum_{i=1}^N \frac{X_{GCE}(V)}{N} = \lim_{V \to \infty} X_{GCE}(V)
\end{equation}
The comparison between (\ref{tlim1}) and (\ref{tlim2}) proves the equivalence
between GCE and CE, MCE. Simplest examples of additive quantities are the energy
and entropy for weakly interacting systems and particle mean multiplicities.

The previous argument does not apply to the variance $\sigma^2$ of particle 
multiplicity distribution. In fact, if we split a CE or MCE into $N$ subsystems,
the variance of any particle multiplicity distribution is not additive, as conservation 
laws involve non-vanishing correlations between different subsystems even for very 
large $N$. So:
\begin{equation}
  \sigma^2 \ne \sum_{i=1}^N \sigma_i^2
\end{equation}
The equal sign would apply only if the subsystems were completely independent
of each other, which is the case only in the GCE. Thus, if variances are not additive,
their GCE and CE thermodynamic limits may and, in general, will differ. Still, the 
variance itself is extensive because the limit (\ref{intensive}) yields a finite 
value in both CE and MCE, as the leading behaviour is ${\cal O}(V)$. 

In conclusion, being non-additive, the variance is a pseudo-extensive quantity and
the scaled variance is thus pseudo-intensive. For such quantities, the thermodynamic
limit in the CE and MCE does not need to coincide with that in the GCE.    

\section{Fluctuations of charged particles}

Fluctuations of charged particle ratios on an event-by-event basis in heavy ion 
collisions have been suggested as probes of the prehadronic phase \cite{jeon,heinz}
and relevant measurements have been performed both at RHIC \cite{star} and 
SPS \cite{na49}. To start with, it is very important to stress that the mere
variance of a ratio of extensive quantities is an ill-defined observable 
in statistical mechanics because it is not a (pseudo-)intensive quantity and 
vanishes in the thermodynamic limit simply because it is proportional to $1/V$. 
Considering for instance the ratio $N_j/N_k$ of particle multiplicities of two 
different species $j$ and $k$ we have:
\begin{equation}\label{ratfluct}
   \left\langle \delta \left(\frac{N_j}{N_k}\right)^2 \right\rangle \simeq
   \frac{1}{\langle N_k \rangle^2} \left( \langle \delta N_j^2 \rangle 
   + \frac{\langle N_j \rangle^2}{\langle N_k \rangle^2} \langle \delta N_k^2 \rangle
   - 2\,{\rm cov} (N_j, N_k) \right)    
\end{equation}
Since $|{\rm cov} (N_j, N_k)| \le \sqrt {\langle \delta N_j^2 \rangle
\langle \delta N_k^2 \rangle}$ and $\langle \delta N_k^2 \rangle \propto V$, one 
is left with an expression which decreases at least proportionally to $1/V$. 
Thus, in order to give a sensible fluctuation measure which does not vanish simply 
because the system gets larger, one should form some truly (pseudo-)intensive variable. 

Many such variables have been proposed to measure charge fluctuations in heavy ion 
collisions (for a review see ref. \cite{volo}), e.g. $D$ \cite{jeon}, $\Phi_Q$ 
\cite{mrow}, and $\nu_{\rm dyn}$ \cite{volo}. Their definitions read:
\begin{eqnarray}\label{definition}
 && D= \langle N_{\rm ch} \rangle \left\langle 
 \delta \left(\frac{N_+}{N_-}\right)^2 \right\rangle \nonumber \\
 &&\Phi_Q = \sqrt{\frac{\langle \Delta Q^2 \rangle}{\langle Q \rangle}}-
 \sqrt{\langle \delta q^2 \rangle}  \\ 
 && \nu_{\rm dyn} = \left\langle \left( \frac{N_+}{\langle N_+ \rangle} - 
 \frac{N_-}{\langle N_- \rangle} \right)^2 \right\rangle - 
 \left( \frac{1}{\langle N_+ \rangle} -\frac{1}{\langle N_- \rangle} \right) 
 \nonumber 
\end{eqnarray}
where $N_+$, $N_-$ and $N_{\rm ch}$ is the number of positive, negative and charged 
particles respectively, $Q= \sum_j Q_j N_j$ is the net charge and $q$ is the charge
of a produced particle. Note that in $\Phi_Q$ definition, in the first term the 
random variables are the numbers $N_j$ of particles with a given charge, whereas
the random variable in the second term is the charge of each particle itself. 
These variables are indeed related to each other \cite{volo} and to the scaled variances
of charged particles. It is not difficult to realize that the first two in 
Eq.~(\ref{definition}) are pseudo-intensive whilst the latter is not and should be 
multiplied by an extensive variable, e.g. $\langle N_{\rm ch}\rangle$ to make it such. 
Being pseudo-intensive, they have different thermodynamic limits in the GCE, CE and
MCE. Therefore, much care is needed in comparing the measurements to the predictions
of statistical mechanics because the effect of conservation laws is crucial in 
determining their values even for very large systems. 

However, this comparison is in general difficult because of additional source of
fluctuations which cannot be disregarded. Even if a statistical model framework would
be essentially correct, there could be large fluctuations of thermodynamic parameters 
(volume, temperature, baryon density etc.) from event to event, which, being 
superimposed to the purely thermodynamic fluctuations, could swamp the thermodynamical
fluctuations. Furthermore, experimental measurements cover a limited kinematical 
window and this introduces a further complication.

\begin{figure}[h]
\begin{center}
  \epsffile{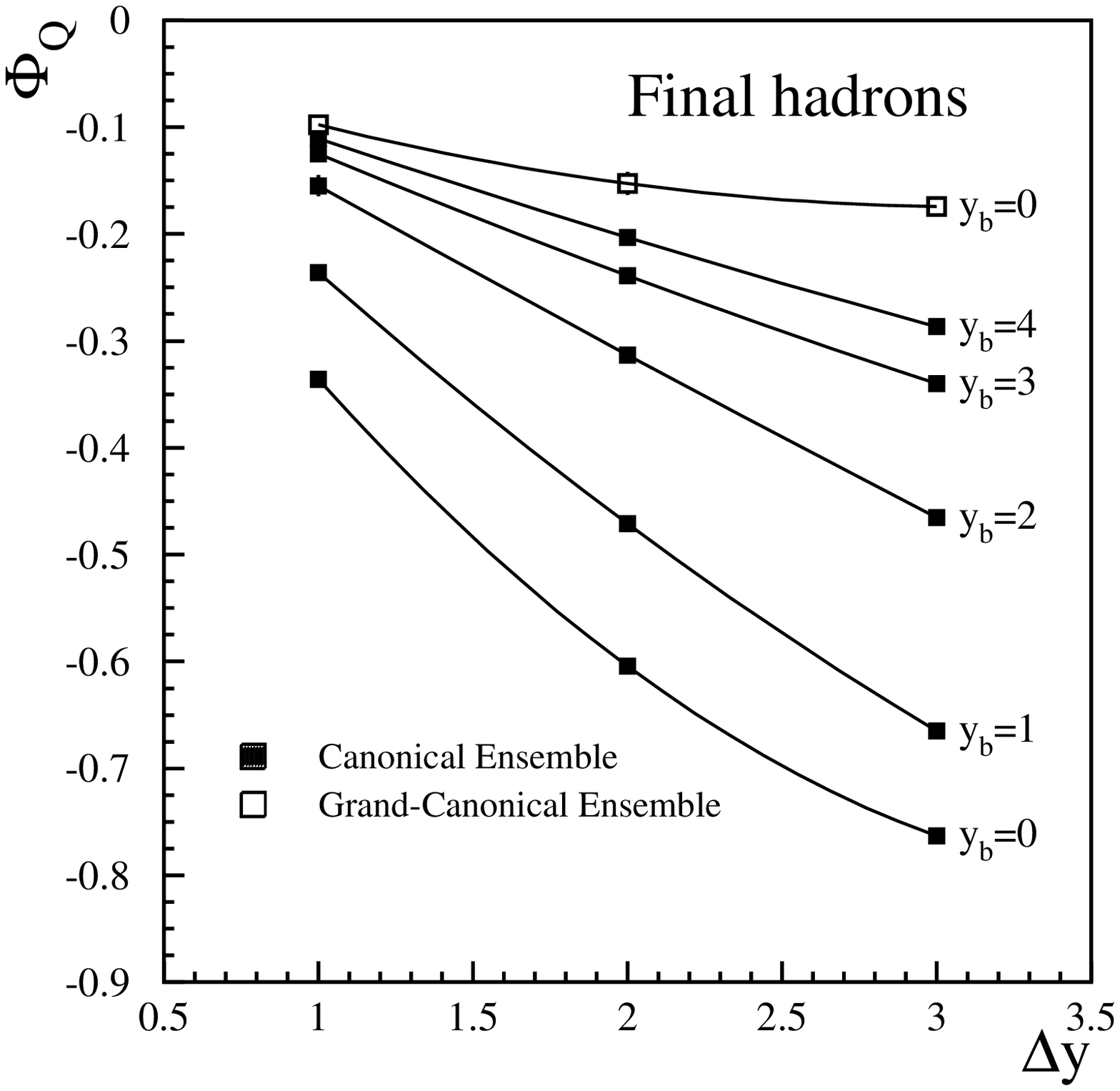} 
\end{center}
\caption{Calculated $\Phi_Q$ (see text for definition) in the canonical and 
grand-canonical ensembles of the full ideal hadron-resonance gas at $T=160$ MeV
and $\rho_B=0.2$ fm$^{-3}$, in the thermodynamic limit, at final hadron 
level for different random boosts of primary particles $y_b$ as a function 
of the acceptance rapidity window $\Delta y$.}
\label{phiq}
\end{figure}

\begin{figure}[h]
\begin{center}
  \epsffile{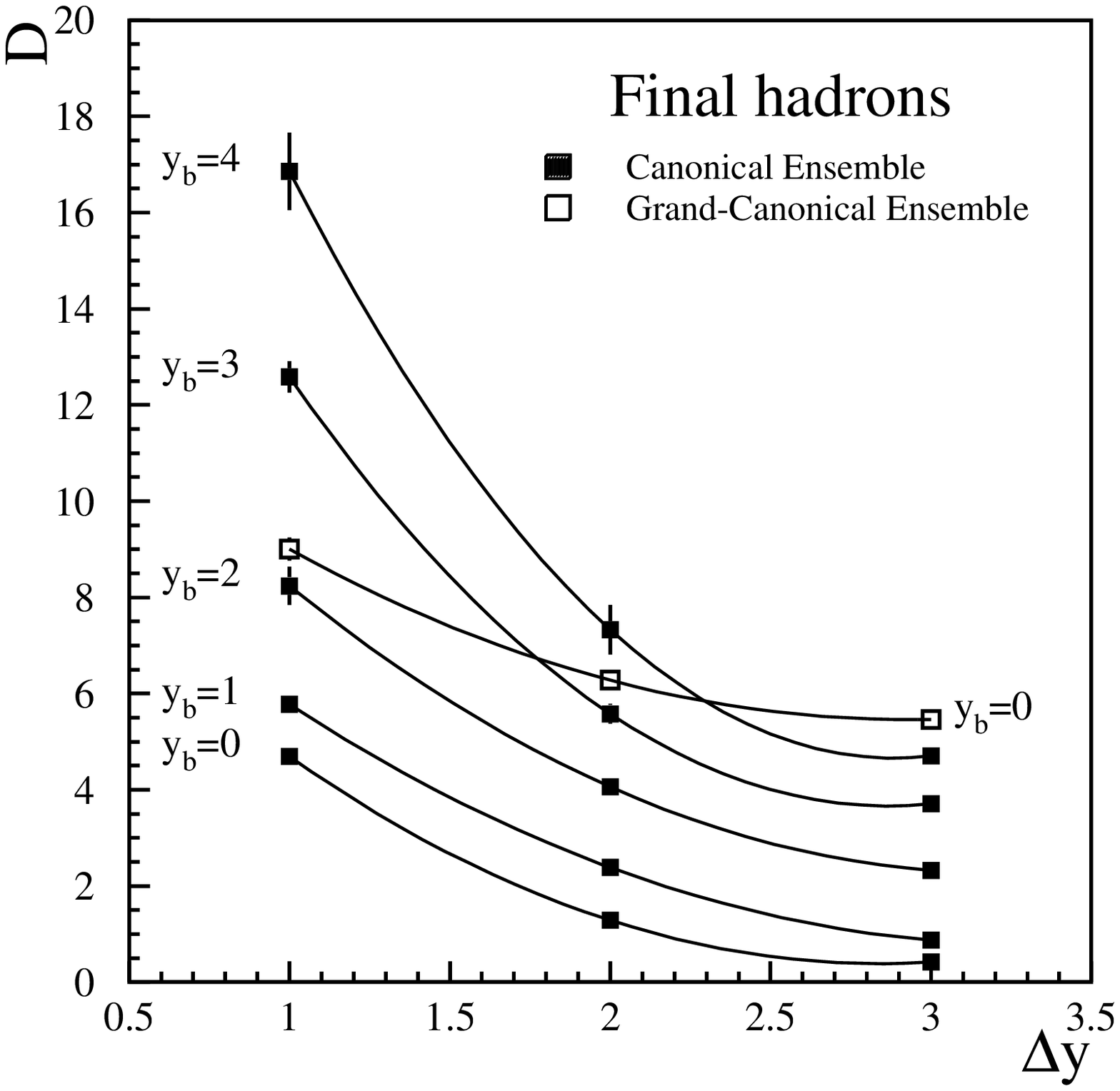} 
\end{center}
\caption{Calculated $D$ (see text for definition) in the canonical and 
grand-canonical ensembles of the full ideal hadron-resonance gas at $T=160$ MeV
and $\rho_B=0.2$ fm$^{-3}$, in the thermodynamic limit, at final hadron 
level for different random boosts of primary particles $y_b$ as a function 
of the acceptance rapidity window $\Delta y$.}
\label{djeon}
\end{figure}

\begin{figure}[h]
\begin{center}
  \epsffile{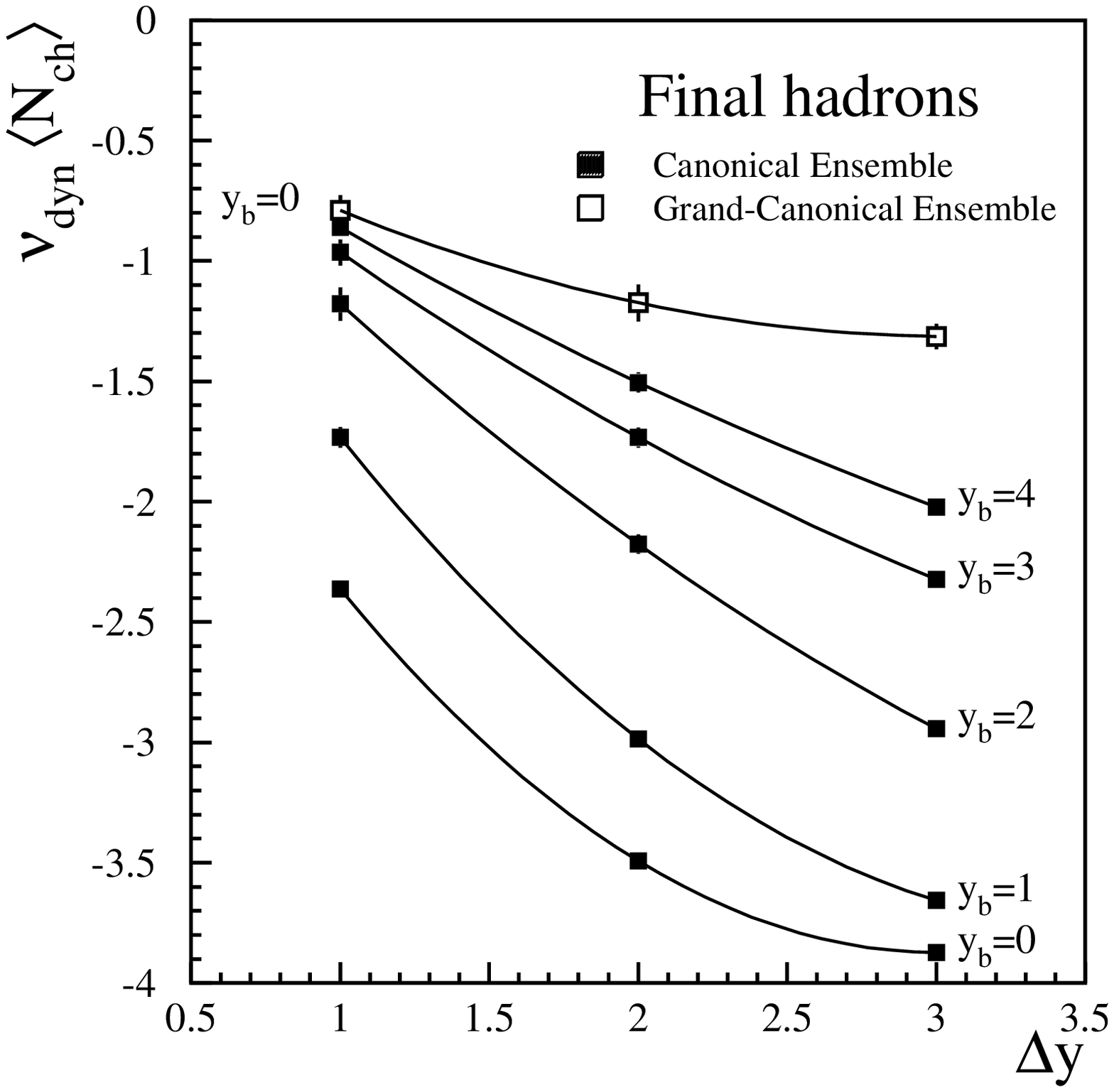} 
\end{center}
\caption{Calculated $\langle N_{\rm ch}\rangle \nu_{\rm dyn}$ (see text for 
definition) in the canonical and grand-canonical ensembles of the full ideal 
hadron-resonance gas at $T=160$ MeV and $\rho_B=0.2$ fm$^{-3}$, in the 
thermodynamic limit, at final hadron level for different random boosts 
of primary particles $y_b$ as a function of the acceptance rapidity window 
$\Delta y$.}
\label{nudyn}
\end{figure}

That said, it can be interesting to study the difference between the fluctuations
of charged particles in the canonical and grand-canonical ensemble. In this respect, 
our Monte-Carlo method is especially suitable as it allows to predict the numerical
values of the aforementioned variables taking into account all ``trivial" effects 
including quantum statistics and resonance decays. Thus, we have calculated the 
values of these variables in the two ensembles at final hadron level for conditions
relevant to heavy ion collisions at $\sqrt s_{NN} \approx 20$ GeV, i.e. $T=160$
MeV, $S=0$, $Q/B=0.4$ and $\rho_B=0.2$ fm$^{-3}$ by using Monte-Carlo simulations.
In the CE, the volume chosen was 200 fm$^3$, which is large enough to ensure in
practice the reaching of the thermodynamic limit. In order to show the effect of
a limited kinematic acceptance, we have also implemented a toy dynamical model, 
giving each primary generated particle (according to thermal distributions) a random 
longitudinal boost in rapidity uniformly between $-y_b$ and $y_b$. Though unrealistic, 
this model allows to understand the possible effect of measuring variables relevant 
to fluctuations over a finite rapidity window $\Delta y$. The calculated values of 
$\Phi_Q$, $D$ and $\langle N_{\rm ch}\rangle \nu_{\rm dyn}$ are shown in 
figs.~\ref{phiq},~\ref{djeon},~\ref{nudyn} respectively. It can be seen 
that all of them are strongly affected by the dynamical boost $y_b$ and the acceptance
window $\Delta y$. Also, a relative strong difference is seen between CE and GCE.
Yet, we note that, at least for $\Phi_Q$ and $\langle N_{\rm ch}\rangle \nu_{\rm dyn}$
their CE value converge to the GC one for small rapidity acceptance. This is
not a trivial feature because such a behaviour is expected if we select a subsystem
in {\em space} and not in {\em momentum space} as we have actually done. In fact,
this behaviour is not seen in $D$ (see fig.~\ref{djeon}). Altogether, we can conclude 
that the spread of these variables is considerable and making a fairly accurate 
estimate of the theoretical expectation for a hadron gas in chemical equilibrium 
requires at least taking into account exact charges (i.e. $B$, $S$, $Q$) conservation.

\section{Summary and conclusions}

We have studied the thermodynamic limits of fluctuations of particle multiplicities
in the full ideal hadron-resonance gas in the canonical and microcanonical ensemble
taking into account the exact conservation of three charges (baryon number, strangeness, 
electric charge) and energy. The inequivalence between these ensembles and the 
grand-canonical ensemble in this respect is clearly seen in our calculations, which 
have been carried out by means of two independent methods: asymptotic expansions 
of the partition functions, giving rise to analytical formulae for scaled variances 
in the thermodynamic limit, and a full Monte-Carlo simulation. Excellent numerical 
agreement has been found between the two. 

The difference between the thermodynamic limits of the scaled variance in the MCE, 
CE and GCE ensembles has its roots in the non-additivity of the variance in the 
former two. In fact, we have shown that all additive quantities, like particle 
multiplicities, must have the same thermodynamic limit. 

We have finally discussed and calculated some variables used to measure charged
particle fluctuations in the canonical and grand-canonical ensemble taking into
account conservation laws and the decay of all resonances. It has been shown that 
their values is, in general, strongly affected by acceptance window and superimposed 
dynamical effects.

\appendix
%
\section*{APPENDIX A  - Saddle-point expansions} 

Generally, we want to find an asymptotic expansion of the complex
$d$-dimensional integral:
\begin{equation}\label{inte}
I(\nu) = \left[\prod_{k=1}^d \frac{1}{2\pi\ii} 
\int_{\Gamma_k} \dii w_k \right] g(\vew) \,  \e^{\nu f(\vew)},
\end{equation}
for large values of the real parameter $\nu$; $\vew$ is a $d$-dimensional vector 
and $g(\vew)$ a smoothly varying function along the paths of integration $\Gamma_k$.
As it is well known, with large values of $\nu$, the dominant contribution to the
integral comes from the small part of the path in the neighbourhood of the saddle 
point $\vew_0$ defined by
\begin{equation}
\vew_0:\ \ \frac{\partial f(\vew)}{\partial w_k}\Bigg|_{\vew_0} = 0 \qquad k=1,\ldots,d.
\end{equation}
In this neighbourhood, one may expand $f(\vew)$ in Taylor series up to second order
as:
\begin{equation}\label{expa}
 f(\vew) \simeq f(\vew_0)+\frac{1}{2} \sum_{i,k} \left(w_i-w_{0i}\right)
  \left(w_k-w_{0k}\right)\frac{\partial^2 f}{\partial w_i \partial w_k}
  \Bigg|_{\vew_0}
\end{equation}
According to the saddle-point or steepest-descent method \cite{web}, we choose a 
real integration variable $t_k$ for the path which is related to the original
$w$ through:
\begin{equation}
  w_k-w_{0k} = \e^{\ii \phi_k} t_k 
\end{equation}
$\phi_k$ being a phase. If $w_{0k}$ lies on the real axis and the eigenvalues of the
Hessian matrix are positive, as it is the case for canonical partition function, 
the steepest descent method prescribes that this phase ought to be $\pi/2$ if the so:
\begin{equation}\label{trans1}
  w_k-w_{0k} = \ii t_k 
\end{equation}
This transformation corresponds to deform the original path into the line 
$w_k = w_{0k}$ in the complex plane. Since only a
small segment around the saddle point contributes to the integral value, we can
extend the integration in $t_k$ from $-\infty$ to $+\infty$ leaving it essentially
unchanged. By plugging the transformation (\ref{trans1}) into Eq.~(\ref{inte}) and 
taking the expansion (\ref{expa}) for $f$, we can write:
\begin{equation}\label{inte2}
  I(\nu) \simeq \e^{\nu f(\vew_0)} \frac{1}{(2\pi)^d} 
  \left[\prod_{k=1}^d \int_{-\infty}^{+\infty} 
  \dii t_k\right] g(\vew(\vet)) \, \e^{-\frac{1}{2} \nu \vet^T {\sf H} \vet}
\end{equation}
where ${\sf H}$ is its Hessian matrix of the function $f(\vew)$ for $\vew=\vew_0$. 
The smooth function $g(\vew)$ is expanded into a Taylor series around the 
saddle-point:
\begin{equation}
g(\vew)=\sum_{n=0}^{\infty} \sum_{\bf n} 
 \frac{\partial^n g(\vew)}{\prod_{k=1}^d \partial w_k^{n_k}} \Bigg|_{\vew_0}
 \prod_{k=1}^d \frac{\left( \ii t_k \right)^{n_k}}{n_k!}
\end{equation}
where ${\bf n} = (n_1,\ldots,n_d)$ labels the sets of $d$ integers
such that $n = \sum_k n_k$. Thus Eq.~(\ref{inte2}) can be rewritten as:
\begin{equation}\label{inte3}
 I(\nu) \simeq \e^{\nu f(\vew_0)} \frac{1}{(2\pi)^d} \sum_{n=0}^{\infty} 
 \ii^n \sum_{\bf n} \frac{\partial^n g(\vew)}{\prod_{k=1}^d \partial w_k^{n_k}} 
 \Bigg|_{\vew_0} \left[\prod_{k=1}^d \frac{1}{n_k!} 
 \int_{-\infty}^{+\infty} \dii t_k \; t_k^{n_k} \right]  
 \e^{-\frac{1}{2} \nu \vet^T {\sf H} \vet}.
\end{equation}
So far, the problem is reduced to solving the integrals:
\begin{equation}\label{single1}
\mathcal{I}_{\bf n} = \left[\prod_{k=1}^d \frac{1}{n_k!} \int_{-\infty}^{+\infty} 
 \dii t_k \; t_k^{n_k} \right] \e^{-\frac{1}{2} \nu \vet^T {\sf H} \vet}
\end{equation}
which is more easily achieved by diagonalizing the symmetric Hessian matrix 
in $\vew_0$, where it is real. Thus, there must be an orthogonal matrix ${\sf A}$
such that:
\begin{equation}
 {\sf H}' = {\mathrm{diag}}(h_1,\ldots,h_d) = {\sf A} {\sf H} {\sf A}^T,
\end{equation}
where $h_k$ are the Hessian's eigenvalues. By defining $\vetau= {\sf A} \vet$,
the integral (\ref{single1}) can be rewritten as:
\begin{equation}\label{single1.5}
\mathcal{I}_{\bf n} = \left[\prod_{k=1}^d \frac{1}{n_k!}
 \int_{-\infty}^{+\infty} 
 \dii \tau_k  \left( \sum_{m=1}^d A_{mk} \tau_m \right)^{n_k} \right]
 \e^{-\frac{1}{2} \nu \vetau^T {\sf H}' \vetau} = 
 \left[\prod_{k=1}^d \frac{1}{n_k!} \int_{-\infty}^{+\infty} 
 \dii \tau_k  \left( \sum_{m=1}^d A_{mk} \tau_m \right)^{n_k} \right]
 \e^{-\frac{1}{2} \nu \sum_{k=1}^d h_k \tau_k^2}
\end{equation}
Now we can use the well known expansion:
\begin{equation}\label{sviluppo}
  \left( \sum_{m=1}^d A_{mk} \tau_m \right)^{n_k} = 
  \sum_{{\bf p}^k(n_k)} \frac{n_k!}{p_1^k!\ldots p_{d}^k!} 
  \prod_{m=1}^d A_{mk}^{p^k_m} \tau_{m}^{p^k_m}
\end{equation}
where ${\bf p}^k(n_k) = (p^k_1,\ldots,p^k_d)$ is a set of $d$ integers such that 
$\sum_{m=1}^d p^k_m = n_k$. Therefore, the integral (\ref{single1.5}) turns to:
\begin{eqnarray}
\mathcal{I}_{\bf n} &=& \prod_{k=1}^d \sum_{{\bf p}^k(n_k)}
\left( \prod_{m=1}^d \frac{A_{mk}^{p^k_m}}{p_m^k!} \right)
 \int_{-\infty}^{+\infty} \dii \tau_k \; \e^{-\frac{1}{2} \nu h_k \tau_k^2} 
 \prod_{m=1}^d \tau_m^{p_m^k} \nonumber \\
  &=& \sum_{{\bf p}^1(n_1),\ldots,{\bf p}^d(n_d)} \prod_{k=1}^d
\left( \prod_{m=1}^d \frac{A_{mk}^{p^k_m}}{p_m^k!} \right)
 \int_{-\infty}^{+\infty} \dii \tau_k \;  
 \tau_k^{p_k^1+\ldots+p_k^d} \e^{-\frac{1}{2} \nu h_k \tau_k^2}
\end{eqnarray}
and, by rescaling the variables $\tau_k \sqrt{h_k \nu} \rightarrow \tau_k$:
\begin{equation}\label{single2}
\mathcal{I}_{\bf n} = \frac{1}{\nu^{d/2}\sqrt{\det{\sf H}}} 
\sum_{{\bf p}^1(n_1),\ldots,{\bf p}^d(n_d)} \prod_{k=1}^d 
 \left(\prod_{m=1}^d \frac{A_{mk}^{p^k_m}}{p_m^k!}    \right)
 \frac{1}{(\nu h_k)^{(p^1_k+\ldots+p^d_k)/2}}
 \int_{-\infty}^{+\infty} \dii \tau_k \tau_k^{p^1_k+\ldots+p^d_k} 
 \e^{-\frac{1}{2}\tau_k^2} 
\end{equation}
Taking into account that $\sum_m p_m^k = n_k$ and $\sum_k n_k = n$ and using known 
formulae for the rightmost gaussian integrals above, we are led to the following
final expression of the integral (\ref{inte3}): 
\begin{equation}\label{intefinal}
 I(\nu) \simeq \e^{\nu f(\vew_0)} \sqrt{\frac{1}{(2\pi \nu)^d \det{\sf H}}} 
 \sum_{n=0}^{\infty} \frac{(-1)^n}{\nu^n} \sum_{\bf n} \frac{\partial^{2n} 
 g(\vew)}{\prod_{k=1}^d \partial w^{n_k}_k} \Bigg|_{\vew_0} \prod_{k=1}^d  
 \sum_{{\bf p}^k(n_k)} \frac{(p^1_k+\ldots+p^d_k-1)!!}{h_k^{(p^1_k+\ldots+p^d_k)/2}} 
 \left(\prod_{m=1}^d \frac{A_{mk}^{p^k_m}}{p_m^k!}\right).
\end{equation} 
where, by convention, $(-1)!!=1$ and $N!!=0$ if $N$ is even. The 
Eq.~(\ref{intefinal}) is admittedly rather cumbersome, but it makes it clear that 
high order terms depend on increasing powers of $1/\nu$, which are negligible for
large values of $\nu$. 

Yet, the expansion (\ref{intefinal}) is not a full asymptotic series for $I(\nu)$ 
because the function $f$ was expanded only at the second order. In order to estimate 
the corrections due to higher order terms in $f$ expansion, one defines a function
$\eta$ such that:
\begin{equation}\label{gamma}
  f(\vew) = f(\vew_0)+\frac{1}{2} \sum_{i,k} \left(w_i-w_{0i}\right)
  \left(w_k-w_{0k}\right)\frac{\partial^2 f}{\partial w_i \partial w_k}
  \Bigg|_{\vew_0} + \eta(\vew)  \; .
  \end{equation}
Obviously:
\begin{equation}\label{gorder}
 \eta(\vew) = {\mathcal O}\left((\vew-\vew_0)^3\right)
\end{equation}
The series (\ref{intefinal}) would turn into an exact asymptotic series provided
that $g$ is replaced with $G(\vew) \equiv g(\vew) \exp[ \nu \eta(\vew)]$. However,
this replacement affects the $\nu$ dependence of the various terms in the series, 
because $G(\vew)$ depends now on $\nu$. This is most easily seen by expanding 
$\exp[\nu \eta(\vew)]$, i.e.:
$$
  G(\vew) = g(\vew) \exp[\nu \eta(\vew)] \simeq g(\vew) + \nu g(\vew) \eta(\vew) 
  + \frac{1}{2} \nu^2 g(\vew) \eta^2(\vew) + \frac{1}{6} \nu^3 g(\vew) \eta^3(\vew)
  + \ldots
$$
Because of (\ref{gorder}), the first two terms in (\ref{intefinal}), with $n=0$ and 
$n=1$, involving derivatives of $G$ of order 0 and 2, are unchanged, while the term 
$n=2$, involving derivatives of $G$ of order 4, is not:
\begin{eqnarray}
  G(\vew_0) &=& g(\vew_0) \nonumber \\
 \frac{\partial^2 G(\vew)}{\prod_{k=1}^d \partial w^{n_k}_k}\Bigg|_{\vew_0} &=&
 \frac{\partial^2 g(\vew)}{\prod_{k=1}^d \partial w^{n_k}_k}\Bigg|_{\vew_0} \nonumber \\
 \frac{\partial^4 G(\vew)}{\prod_{k=1}^d \partial w^{n_k}_k}\Bigg|_{\vew_0} &=&
 \frac{\partial^4 g(\vew)}{\prod_{k=1}^d \partial w^{n_k}_k}\Bigg|_{\vew_0} +  
 \nu g(\vew_0) \frac{\partial^4 f(\vew)}
 {\prod_{k=1}^d \partial w^{n_k}_k}\Bigg|_{\vew_0} + \nu
 \sum_{i=1}^d n_i \frac{\partial g(\vew)}{\partial w_i}\Bigg|_{\vew_0} 
 \frac{\partial^3 f(\vew)]}{\partial w_1^{n_1} \ldots 
 \partial w_{i}^{n_i-1} \ldots \partial w_d^{n_d}}\Bigg|_{\vew_0}
\end{eqnarray}
where we have used Eq.~(\ref{gamma}). 
It can be seen that the rightmost two terms in the last equation are proportional 
to $\nu$, hence they introduce in the term with $n=2$ of the series in 
Eq.~(\ref{intefinal}) quantities of the order ${\mathcal O}(\nu^{-1})$. Furthermore, 
even in the term $n=3$ of the series a term in the sixth derivatives of $G$ appears 
which turns out to be of the order ${\mathcal O}(\nu^{-1})$, that is:
\begin{equation}
 \frac{1}{2} \nu^2 g(\vew_0)  
 \frac{\partial^6 f(\vew)}{\prod_{k=1}^d \partial w^{n_k}_k}\Bigg|_{\vew_0}
\end{equation}
We can finally write the full expansion of $I(\nu)$ at order ${\mathcal O}(\nu^{-1})$
as:
\begin{eqnarray}
 I(\nu) &\simeq& \e^{\nu f(\vew_0)} \sqrt{\frac{1}{(2\pi\nu)^d \det{\sf H}}}
 \left\{ g(\vew_0) + \frac{1}{\nu} \left[ - \frac{1}{2} \sum_{k,m=1}^d \right. \right.
 \frac{\partial^2 g(\vew)}{\partial w_k \partial w_m} \Bigg|_{\vew_0} 
 \left( \sum_{i=1}^d \frac{A_{im} A_{ik}}{h_i} \right) \nonumber \\
 &+& g(\vew_0) \sum_{{\bf n},|{\bf n}|=4} \frac{\partial^4 f(\vew)}
 {\prod_{k=1}^d \partial w^{n_k}_k} \Bigg|_{\vew_0} \prod_{k=1}^d  
 \sum_{{\bf p}^k(n_k)} \frac{(p^1_k+\ldots+p^d_k-1)!!}{h_k^{(p^1_k+\ldots+p^d_k)/2}} 
 \left(\prod_{m=1}^d \frac{A_{mk}^{p^k_m}}{p_m^k!}\right) \nonumber \\
 &+& \sum_{{\bf n},|{\bf n}|=4} 
 \sum_{i=1}^d n_i \frac{\partial g(\vew)}{\partial w_i}\Bigg|_{\vew_0} 
 \frac{\partial^3 f(\vew)]}{\partial w_1^{n_1} \ldots 
 \partial w_{i}^{n_i-1} \ldots \partial w_d^{n_d}}\Bigg|_{\vew_0}
 \prod_{k=1}^d 
 \sum_{{\bf p}^k(n_k)} \frac{(p^1_k+\ldots+p^d_k-1)!!}{h_k^{(p^1_k+\ldots+p^d_k)/2}} 
 \left(\prod_{m=1}^d \frac{A_{mk}^{p^k_m}}{p_m^k!}\right) \nonumber \\
 &-& \frac{1}{2} g(\vew_0) \sum_{{\bf n},|{\bf n}|=6} \frac{\partial^{6} f(\vew)}
 {\prod_{k=1}^d \partial w^{n_k}_k} \Bigg|_{\vew_0} \prod_{k=1}^d 
 \sum_{{\bf p}^k(n_k)} \frac{(p^1_k+\ldots+p^d_k-1)!!}{h_k^{(p^1_k+\ldots+p^d_k)/2}} 
 \left.\left. \left(\prod_{m=1}^d \frac{A_{mk}^{p^k_m}}{p_m^k!} 
 \right) \right] \right\}.
\end{eqnarray}
where, by convention $|{\bf n}| = \sum_k n_k$. The above expression can be shortly 
summarized as:
\begin{equation}\label{core}
 I(\nu) \simeq \e^{\nu f(\vew_0)} \sqrt{\frac{1}{(2\pi\nu)^d \det{\sf H}}}
 \left\{ g(\vew_0) + \frac{1}{\nu} \left[ - \frac{1}{2} \sum_{k,m=1}^d 
 \frac{\partial^2 g(\vew)}{\partial w_k \partial w_m} \Bigg|_{\vew_0} 
 \left( \sum_{i=1}^d \frac{A_{im} A_{ik}}{h_i} \right) 
 + \sum_{i=1}^d \alpha_i \frac{\partial g(\vew)}{\partial w_i}\Bigg|_{\vew_0} 
 + \gamma g(\vew_0) \right] \right\}
\end{equation}
with $\gamma$ and $\alpha$ being constants dependent {\em only} on the function 
$f$ and its derivatives.

\section*{APPENDIX B  - Saddle-point expansion of the microcanonical 
partition function} 

To work out the asymptotic expansion of the MPF in Eq.~(\ref{mpf}) we define our
$f$ and $g$ functions like in Appendix A as:
\begin{eqnarray}\label{gf}
  g(\vew) &=& w_E^{-1} w_B^{-1} w_S^{-1} w_Q^{-1} \\
  \label{microf}
  f(\vew) &=& \rho_E \ln w_E - \rho_B \ln w_B - \rho_S \ln w_S - \rho_Q \ln w_Q + 
  \sum_j F^{(0)}_j(w_E) w_B^{b_j}w_S^{s_j}w_Q^{q_j}
\end{eqnarray}
Thus:
\begin{equation}
  \Omega_{\veQ} = \lim_{\epsilon \to 0^+} \frac{1}{(2\pi\ii)^4}
 \int_{-\ii \infty+ \epsilon}^{+\ii \infty+\epsilon} \dii w_E\
  \oint \dii w_B \oint \dii w_S \oint \dii w_Q\ \; g(\vew) \exp[V f(\vew)]
\end{equation}
By using the expansion~(\ref{core}) at the leading order for the functions 
$\Omega_{\veQ}$ and $\Omega_{\veQ}^{(j)}$ defined in Eq.~(\ref{omegaj}), plugging
in the saddle point equations (\ref{saddle2}) defining temperature and fugacities,
and proceeding like in  Subsect.~\ref{subs:general}, it is rather straightforward 
to obtain, in the limit of Boltzmann statistics:
\begin{equation} 
 \lim_{V \to \infty} \langle N_h \rangle = \lim_{V \to \infty} 
 V \sum_{j\in h} \frac{\Omj_{\veQ-\veq_{j}}}{\Omega_{\veQ}} =
 \sum_{j\in h} z_j \lambda_j 
\end{equation} 
with $\lambda_j = \lambda_B^{b_j} \lambda_S^{s_j} \lambda_Q^{q_j}$, 
that is the thermodynamic limit of mean multiplicities in the MCE is the
corresponding value in the GCE.

Let us now turn to scaled variances. Denoting $\hF^{(n)}_j\equiv F^{(n)}_j(\e^\beta)$,
with $F^{(n)}_j(w_E)$ given by Eq.~(\ref{ffunct}):
$$
 F^{(n)}_j(w_E) \equiv \frac{(2J_j+1)}{(2\pi)^3} \int \dii^3 p \; 
 \varepsilon_j^n w_{E}^{-\varepsilon_j}
$$
and $\vew_0 = (\e^\beta,\Lab,\Las,\Laq)$, the Hessian of $f$ reads:
\begin{eqnarray}
  H_{00}(\vew_0)&=& \frac{1}{V} \, \e^{-2\beta} \sum_j \hF_j^{(2)} 
  \lambda_j \nonumber \\
  H_{0l}(\vew_0)  &=& - \frac{1}{V \lambda_{Q_l}} \, \e^{-\beta} 
  \sum_j q_{l,j} \hF_j^{(1)}\lambda_j        \nonumber \\
  H_{ln}(\vew_0) &=& \frac{1}{V \lambda_{Q_l}\lambda_{Q_n}} 
 \left[ Q_l \delta_{ln} + \sum_j q_{l,j} \left( q_{n,j}-\delta_{ln}\right)
 \langle N_j \rangle_{\GC} \right] \nonumber
\end{eqnarray}
If we now take the general expression of the asymptotic expansion in (\ref{core}),
it can be seen that the function $g$ enters with both first and second order
derivatives. However, since we are interested in calculating scaled variances,
only the term involving second derivatives yields a non-vanishing contribution,
just like in the canonical case. We will then confine ourselves to calculate
the latter, taking into account that $g$ varies according to which $\Omega$-related
quantity we are interested in. In fact:
\begin{eqnarray}\label{g0}
    \Omega_{\veQ} &\rightarrow&  g(\vew) = w_E^{-1} w_B^{-1} w_S^{-1} w_Q^{-1}
    \\
 \label{g1}
    \Omj_{\veQ-\veq_{j}} &\rightarrow&  g^{(j)}(\vew) = F^{(0)}_j 
    w_E^{-1} w_B^{b_j-1} w_S^{s_j-1} w_Q^{q_j-1}
  \\
 \label{g2} 
    \Omjk_{\veQ-\veq_{j}-\veq_{k}} &\rightarrow&  g^{(jk)}(\vew) = 
    F^{(0)}_j F^{(0)}_k w_E^{-1} w_B^{b_j+b_k-1} w_S^{s_j+s_k-1} w_Q^{q_j+q_k-1}
\end{eqnarray}
Nowe we can calculate the second derivatives of the functions $g$. Taking into
account that:
\begin{equation}
  \frac{\partial F^{(n)}_j(\vew)}{\partial w_l} = -\frac{F^{(n+1)}_j(\vew)}{w_l}
\end{equation}
it can be easily shown that
\begin{equation}\label{zero}
  \left.\frac{\partial^2 g(\vew)}{\partial w_l \partial w_n}\right|_{\vew_0}
  = (1+\delta_{ln}) (w_0)_l^{-1}(w_0)_n^{-1}\prod_{m=0}^{3}(w_0)_m^{-1}  
\end{equation}
for Eq.~(\ref{g0}),
\begin{eqnarray}\label{one}
\left.\frac{\partial^2 g^{(j)}(\vew)}{\partial w_E^2}\right|_{\vew_0}
&=& \e^{-\beta} \lambda_B^{b_j-1}\lambda_S^{s_j-1}\lambda_Q^{q_j-1}
\left(2\hF_j^{(0)}+3\hF_j^{(1)}+\hF_j^{(2)}\right) \e^{-2\beta}
\\
\left.\frac{\partial^2 g^{(j)}(\vew)}{\partial w_E\partial w_{Q_l}}
\right|_{\vew_0} 
&=& -\e^{-\beta} \lambda_B^{b_j-1}\lambda_S^{s_j-1}\lambda_Q^{q_j-1}
\left( q_{l,j}-1 \right) \left( \hF_j^{(0)}+\hF_j^{(1)}\right) \e^{-\beta}
\lambda_{Q_l}^{-1}
\nonumber \\
\left.\frac{\partial^2 g^{(j)}(\vew)}{\partial w_{Q_l}\partial w_{Q_n}}
\right|_{\vew_0}
&=& \e^{-\beta} \lambda_B^{b_j-1}\lambda_S^{s_j-1}\lambda_Q^{q_j-1}
\left( q_{l,j}-1 \right) \left(q_{n,j}-1-\delta_{ln}\right) \hF_j^{(0)}
\lambda_{Q_l}^{-1}\lambda_{Q_n}^{-1}\nonumber
\end{eqnarray}
for Eq.~(\ref{g1}), and
\begin{eqnarray}\label{two}
\left.\frac{\partial^2 g^{(jk)}(\vew)}{\partial w_E^2}\right|_{\vew_0}
&=& \e^{-\beta} \lambda_B^{b_j+b_k-1}\lambda_S^{s_j+s_k-1}\lambda_Q^{q_j+q_k-1}
\left(
2\hF_j^{(0)}\hF_k^{(0)}+3\hF_j^{(1)}\hF_k^{(0)}+3\hF_j^{(0)}\hF_k^{(1)}\right.\\
&&\left.+2\hF_j^{(1)}\hF_k^{(1)}+\hF_j^{(2)}\hF_j^{(0)}+\hF_j^{(0)}\hF_k^{(2)}\right) 
\e^{-2\beta} 
\nonumber\\
\left.\frac{\partial^2 g^{(jk)}(\vew)}{\partial w_E \partial w_{Q_l}}
\right|_{\vew_0} 
&=& -\e^{-\beta} \lambda_B^{b_j+b_k-1}\lambda_S^{s_j+s_k-1}\lambda_Q^{q_j+q_k-1}
\left(q_{l,j}+q_{l,k}-1\right) \left(\hF_j^{(0)}\hF_k^{(0)}+\hF_j^{(1)}
\hF_k^{(0)} + \hF_j^{(0)}\hF_k^{(1)}\right) \e^{-\beta}\lambda_{Q_l}^{-1}
\nonumber \\
\left.\frac{\partial^2 g^{(jk)}(\vew)}{\partial w_{Q_l}\partial w_{Q_n}}
\right|_{\vew_0}
&=& \e^{-\beta} \lambda_B^{b_j+b_k-1}\lambda_S^{s_j+s_k-1}\lambda_Q^{q_j+q_k-1}
\left(q_{l,j}+q_{l,k}-1\right) \left(q_{n,j}+q_{n,k}-1-\delta_{ln}\right)
\hF_j^{(0)} \hF_k^{(0)} \lambda_{Q_l}^{-1} \lambda_{Q_n}^{-1}.\nonumber
\end{eqnarray}
for Eq.~(\ref{g2}).

We can now use the general expression (\ref{core}) obtained in Appendix A to
express the factor in parentheses in Eq.~(\ref{mcomega}). Just like in the 
canonical ensemble case, the terms proportional to $g$ and to the first derivatives 
of $g$ within brackets in Eq.~(\ref{core}) with coefficients $\alpha$ and $\gamma$,
do not contribute to the scaled variance in the limit $V \to \infty$. Indeed,
defining:
\begin{equation}
  M_{ln} = \sum_{i=0}^3 \frac{A_{il} A_{in}}{h_i} 
  \qquad{\mathrm {and}}\qquad
  \frac{\partial^2 g}{\partial w_l \partial w_m} = g_{l,m}
\end{equation}
we obtain, at the order ${\mathcal O}(V^{-1})$:
\begin{equation}\label{omcoeff}
 \frac{\Omjk_{\veQ-\veq_k-\veq_j}}{\Omj_{\veQ-\veq_j}} - 
 \frac{\Omk_{\veQ-\veq_k}}{\Omega_{\vec{Q}}} \simeq 
  \frac{1}{V} \sum_{l,n=1}^d M_{ln} 
  \left[ \frac{g^{(k)}(\vew_0)g_{l,n}(\vew_0)}{g(\vew_0)^2} 
         -\frac{g^{(jk)}(\vew_0)g_{l,n}^{(j)}(\vew_0)}{g^{(j)}(\vew_0)^2}
	 -\frac{g_{l,n}^{(k)}(\vew_0)}{g(\vew_0)}
	 +\frac{g_{l,n}^{(jk)}(\vew_0)}{g^{(k)}(\vew_0)}
  \right]
\end{equation}
Using Eqs.~(\ref{zero}),(\ref{one}) and (\ref{two}) to work out the 
expression within brackets in Eq.~(\ref{omcoeff}) we finally obtain the 
scaled variance in the thermodynamic limit of MCE quoted in Eq.~(\ref{mctl}):
\begin{eqnarray}
\lim_{V \to \infty} \omega_h = 1 &-& \frac{1}{\sum_{j\in h} \langle N_j \ranGC}
\sum_{j\in h} \langle N_j \ranGC \sum_{k\in h} \lambda_k 
\left\{ \frac{\hF_j^{(1)}\hF_k^{(1)}}{\hF_j^{(0)}}\e^{-2\beta} M_{00}\right. 
\nonumber\\
&& -\sum_{l=1}^3 \left[q_{l,k} \hF_j^{(1)}\hF_k^{(0)}+q_{l,j} \hF_j^{(0)}\hF_k^{(1)} 
\right] \frac{\e^{-\beta} \lambda_{Q_l}^{-1}} {\hF_j^{(0)}} M_{0l} 
\left. + \sum_{l,n=1}^3 q_{l,j}q_{m,k} \hF_k^{(0)}\lambda_{Q_l}^{-1}
 \lambda_{Q_n}^{-1} M_{ln}\right\}.
\end{eqnarray}
%

\begin{acknowledgments}
We gratefully acknowledge useful discussions with M. Gazdzicki, 
M. Gorenstein, K. Rajagopal, M. Stephanov, H. Touchette.
\end{acknowledgments}




\pagebreak


\end{document}